\documentclass[acmsmall]{acmart}
\AtBeginDocument{%
  }

\setcopyright{cc}
\setcctype{by-nc-nd}
\acmDOI{10.1145/3808193}
\acmYear{2026}
\acmJournal{PACMSE}
\acmVolume{3}
\acmNumber{FSE}
\acmArticle{FSE186}
\acmMonth{7}
\acmSubmissionID{fse26mainb-p2166-p}
\received{2026-02-21}
\received[accepted]{2026-03-24}

\usepackage{graphicx} %
\usepackage{color}
\usepackage{subcaption}
\usepackage{caption}
\usepackage{cleveref}
\usepackage{amsmath}
\usepackage{multirow}
\usepackage{algorithm}
\usepackage{algorithmic}
\usepackage{xspace}
\usepackage{makecell}
\usepackage{pifont}
\usepackage{pgfplots}
\usepackage{wrapfig}
\usepackage{hyperref}
\usepackage[most]{tcolorbox}
\usepackage{enumitem}
\setlist[itemize]{leftmargin=*}

\newcommand{\bella}[1]{}
\newcommand{\reply}[1]{}

\newcommand{\rqbox}[1]{

\begin{tcolorbox}[tile, size=fbox, boxsep=2mm, boxrule=0pt, top=0pt, bottom=0pt,
borderline west={1mm}{0pt}{blue!50!white}, colback=blue!5!white]
#1
\end{tcolorbox}

}

\newtcolorbox[auto counter, number within=section]{promptbox}[1][]{
    colback=black!3!white,
    colframe=black!75!white,
    fonttitle=\bfseries,
    breakable,              %
    title={Prompt \thetcbcounter: #1}
}

\begin{document}

\title{\textsc{iCoRe}: An Iterative Correlation-Aware Retriever for Bug Reproduction Test Generation}

\author{Junyi Wang}
\orcid{0009-0001-7058-9071}
\email{wangjunyi@zju.edu.cn}
\affiliation{%
  \department{College of Computer Science and Technology}
  \department{The State Key Laboratory of Blockchain and Data Security}
  \institution{Zhejiang University}
  \city{Hangzhou}
  \country{China}
}

\author{Jialun Cao}
\orcid{0000-0003-4892-6294}
\email{jialuncao@ust.hk}
\affiliation{%
  \institution{The Hong Kong University of Science and Technology}
  \city{Hong Kong}
  \country{China}
}

\author{Zhongxin Liu}
\authornote{Corresponding author. Also with Hangzhou High-Tech Zone (Binjiang) Institute of Blockchain and Data Security.}
\orcid{0000-0002-1981-1626}
\email{liu\_zx@zju.edu.cn}
\affiliation{%
  \department{College of Computer Science and Technology}
  \department{The State Key Laboratory of Blockchain and Data Security}
  \institution{Zhejiang University}
  \city{Hangzhou}
  \country{China}
}

\renewcommand{\shortauthors}{Junyi Wang, Jialun Cao, Zhongxin Liu}
\newcommand{\appname}{{\sc iCoRe}\xspace}

\begin{abstract}
    Automatically generating bug reproduction tests (BRT) from issue descriptions is crucial for facilitating software maintenance. 
    Large Language Model (LLM)-based approaches have shown great potential for this task. 
    Their effectiveness heavily relies on retrieving high-quality context from the codebase.
    The retrieval phase of existing approaches relies on either traditional methods like BM25 or modern LLM-driven strategies. 
    The LLM-based retrieval strategies typically involve equipping an LLM with tools to autonomously explore the code repository or having it select the most relevant files and code snippets from a provided list as context.
    However, these retrieval methods suffer from three key limitations: 
    (1) They often employ a unified strategy for retrieving both source code and test cases, overlooking their distinct retrieval requirements. 
    (2) They focus solely on semantic similarity, ignoring function call relationships that reflect behavioral relevance, which often leads to the retrieval of irrelevant context. 
    (3) The retrieval lacks a feedback loop from the generation phase, preventing it from refining the context based on execution
    results. 
    These limitations collectively result in low-quality context, thereby hindering the accuracy of bug reproduction.
    
    To address these challenges, we propose \appname
    , an iterative, correlation-aware context retrieval approach. 
    \appname is explicitly designed to be aware of three key correlations: 1) the correlation between source code and test cases, which requires differentiated retrieval, 2) the correlation between textual semantics and function call structures for accurate relevance assessment, and 3) the correlation between the retrieval and generation phases, which enables iterative feedback and refinement.
        To evaluate \appname, we integrate it with an LLM-based BRT generator and conduct a comprehensive evaluation on the SWT-bench Lite and TDD-bench Verified benchmarks.
        Experimental results show that our method achieves a Fail-to-Pass rate of 42.0\% and 52.8\% respectively, representing significant 19.7\%--31.7\% relative improvements over existing retrieval methods.
\end{abstract}



\maketitle

\section{Introduction}

Bug reproduction refers to the process of reliably triggering the faulty behavior of a reported software issue, enabling developers to observe, analyze, and ultimately diagnose the underlying defect.
In a typical software development workflow, once a bug is reported, a developer's task is to first reproduce the bug, then locate the root cause of the problem, and finally write a fix patch to resolve it~\cite{10.1007/978-3-319-06498-7_12}.
The executable test created during this process to reproduce the bug is known as a Bug Reproduction Test (BRT).
Given an issue description and its associated buggy code repository, a valid BRT should satisfy two core conditions:
(1) it should fail on the buggy version of the code, thereby manifesting the undesired behavior;
(2) it should succeed after the fix patch is applied, thereby validating the correctness of the applied fix \cite{cheng2025agenticbugreproductioneffective,mundler2024swtbench}.

Well-constructed Bug Reproduction Tests (BRTs) are crucial for software maintenance, as they help developers locate issues, validate fixes, and prevent regressions~\cite{10.1007/978-3-319-06498-7_12,6982627,8804445}.
However, manually creating these tests is a time-consuming and labor-intensive process ~\cite{10.1145/3468264.3473922,chen2024b4,liu2023towards}.
To address this, the automated generation of BRTs has become a significant area of research
~\cite{10.1007/978-3-319-99241-9_18,7081820jcharming,libro_kang_yoon_yoo_2023}, which not only enhances the efficiency of human developers but also improves the performance of automated issue-solving agents ~\cite{arora2024masaimodulararchitecturesoftwareengineering,sweagent_yang_et_al_2024,wang2025openhands}.
For human developers, this automation significantly reduces the burden of manual bug reproduction, allowing them to focus their efforts directly on root cause analysis and implementing fixes
~\cite{8453125,cheng2025agenticbugreproductioneffective,aegis}.
For automated issue-solving agents, many of them incorporate BRT generation as a key component, relying on the resulting tests to validate whether their proposed fixes have correctly resolved the bug~\cite{10.1007/s10664-026-10802-w,coder_chen_et_al_2024,xia2025demystifying,autocoderover_zhang_et_al_2024}.

Traditional methods often rely on analyzing specific program crash data like stack traces~\cite{6926857,9286108}, thus are ill-suited for bugs described in natural language.
Recent advances in Large Language Models (LLMs) have established the new state-of-the-art in automated BRT generation, thanks to their strong capabilities in natural language understanding and code generation.
Existing LLM-based methods can be divided into two types. 
Some methods~\cite{libro_kang_yoon_yoo_2023,mundler2024swtbench} operate by directly prompting an LLM with few-shot examples or context retrieved via BM25.
Other methods~\cite{aegis,otter_ahmed_et_al_2025,ahmed2025executionfeedbackdriventestgeneration,khatib2025assertflipreproducingbugsinversion} adopt a more complex, multi-step process: they first employ an LLM-based retriever to localize relevant files and code snippets, and then initiate an iterative cycle of generation, execution, and refinement.

Accurately retrieving relevant context is crucial for the success of LLM-based bug reproduction.
issue descriptions frequently lack the key details required for reproducing bugs  ~\cite{10.1145/1453101.1453146,Rahman_2020,dou2024s}.
Without sufficient context, an LLM is prone to making errors for bug reproduction~\cite{liu2024exploringevaluatinghallucinationsllmpowered}. 
For example, it may use the wrong testing framework or misuse function APIs. 
In contrast, accurate context equips the model with key information like correct API usage and example test cases, significantly reducing hallucinations and improving generation quality ~\cite{10.1145/3728894,10.1145/3660783,zhang2025unit}.
This necessity is also confirmed by recent studies. 
For instance, removing context localization causes a 17.7\% performance drop for the LLM-based BRT generation approach Otter ~\cite{otter_ahmed_et_al_2025}, and for its state-of-the-art successor e-Otter, the success rate was 42.6\% among instances with correct test localization, dropping sharply to 24.2\% for those with incorrect localization ~\cite{ahmed2025executionfeedbackdriventestgeneration}.

Existing LLM-based bug reproduction approaches have adopted a variety of retrieval strategies to provide contextual information.
Some methods rely on traditional term-based retrieval like BM25\cite{mundler2024swtbench}, while more recent approaches use Large Language Models to identify relevant context.
This is typically done either by equipping the LLM with tools as a search agent to explore the code repository autonomously~\cite{aegis}, or by having it perform a hierarchical selection, i.e., first choosing relevant files from a list, and then identifying relevant functions within them \cite{khatib2025assertflipreproducingbugsinversion,otter_ahmed_et_al_2025,ahmed2025executionfeedbackdriventestgeneration}.
However, these retrievers face the following key limitations:

\textbf{Absence of Differentiation Between Production Code and Test Code Retrieval.}
Effective bug reproduction requires two distinct types of context: production code and existing tests, each with unique retrieval needs. 
Production code,
which clarifies API usage, is typically found by matching explicit entities from a issue description, such as function names. 
In contrast, existing tests, which provide vital structural templates, require understanding the described behavioral scenario.
However, the retrieval phase of existing approaches often employs a single, unified strategy for both, ignoring their distinct requirements.
For example, AssertFlip \cite{khatib2025assertflipreproducingbugsinversion} retrieves only potentially faulty production code and completely ignores the vital test templates offered by existing tests. 
Ultimately, this absence of a differentiated strategy results in an incomplete context, thereby hindering the successful reproduction of bugs. 
    
\textbf{Neglect of Function Call Relationships.} 
In this work, we define Function Call Relationships as the structural connections between code entities, such as functions, methods, and classes, formed by caller-callee interactions.
Both traditional BM25 and modern LLM-driven retrievers ignore this crucial structural information.
BM25 relies on term matching, meaning it can only identify shared keywords between a issue description and a piece of code.
When LLM-driven strategies analyze a file or code snippet, it determines relevance by understanding the textual meaning and its similarity to the bug description. 
Consequently, both approaches are prone to retrieving code and tests that are textually or semantically aligned with the issue description but behaviorally irrelevant.
For example, as illustrated in Figure \ref{fig:otter_example}, the LLM-based retrieval module of Otter \cite{otter_ahmed_et_al_2025} retrieves test cases based on semantic similarity while neglecting function call relationships.
Consequently, it retrieves the test case ``test\_combine\_negated\_boolean\_expression'', whose implementation does not invoke the key functions, is irrelevant, and did not help to reproduce the bug.
    
\textbf{Lack of Feedback from Generation to Retrieval.} 
Existing BRT generation approaches treat retrieval as an isolated step performed before generation.
This ``one-way'' workflow completely overlooks the valuable information from the generation phase.
By leveraging feedback from the generation phase, such as the textual content and function call structure of the generated test, the retrieval process can be refined to identify more relevant context, thereby enhancing bug reproduction performance.
For instance, as illustrated in Figure \ref{fig:aegis_example}, an agent might find a suboptimal test like ``test\_filtering\_on\_annotate\_that\_uses\_q'' and stop. 
However, the test generated from this context provides valuable feedback that could guide a subsequent retrieval round to the initially missed but helpful context ``test\_empty\_expression\_annotation''.

\vspace{-0.1cm}
\begin{figure}[htbp]
  \centering
  \begin{subfigure}{\linewidth}
    \includegraphics[width=\linewidth]{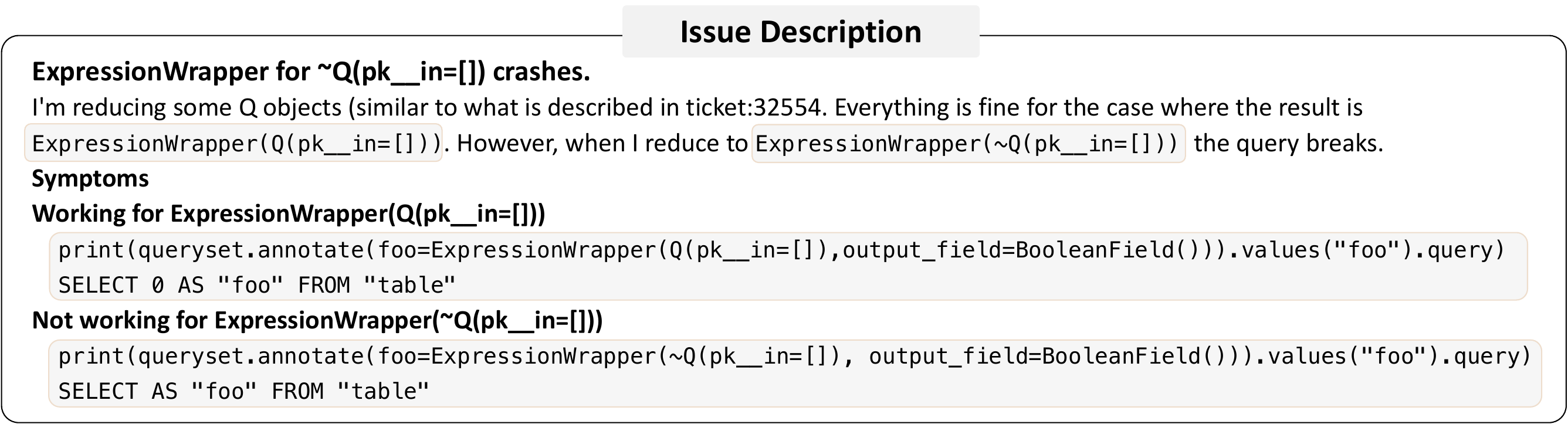}
    \caption{The issue description of ``django\_\_django-15213.''}
    \label{fig:prob_decs}
  \end{subfigure}
  \hfill
  \begin{subfigure}{\linewidth}
    \includegraphics[width=\linewidth]{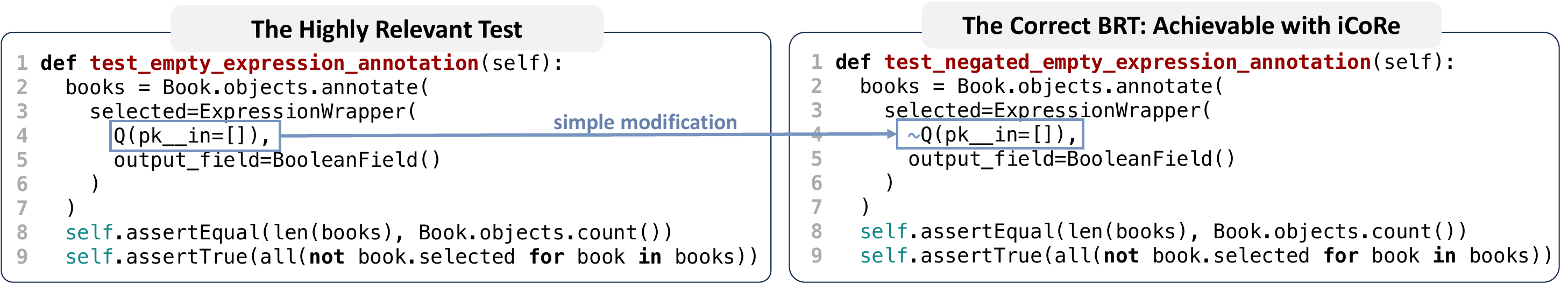}
    \caption{An straightforward method for reproducing ``django\_\_django-15213.''}
    \label{fig:solution}
  \end{subfigure}
  
  \caption{Motivating Example}
  \label{fig:motivating_example}
\end{figure}

\begin{figure}[htbp]
  \centering
  \begin{subfigure}{0.48\linewidth}
    \centering
    \includegraphics[width=\linewidth]{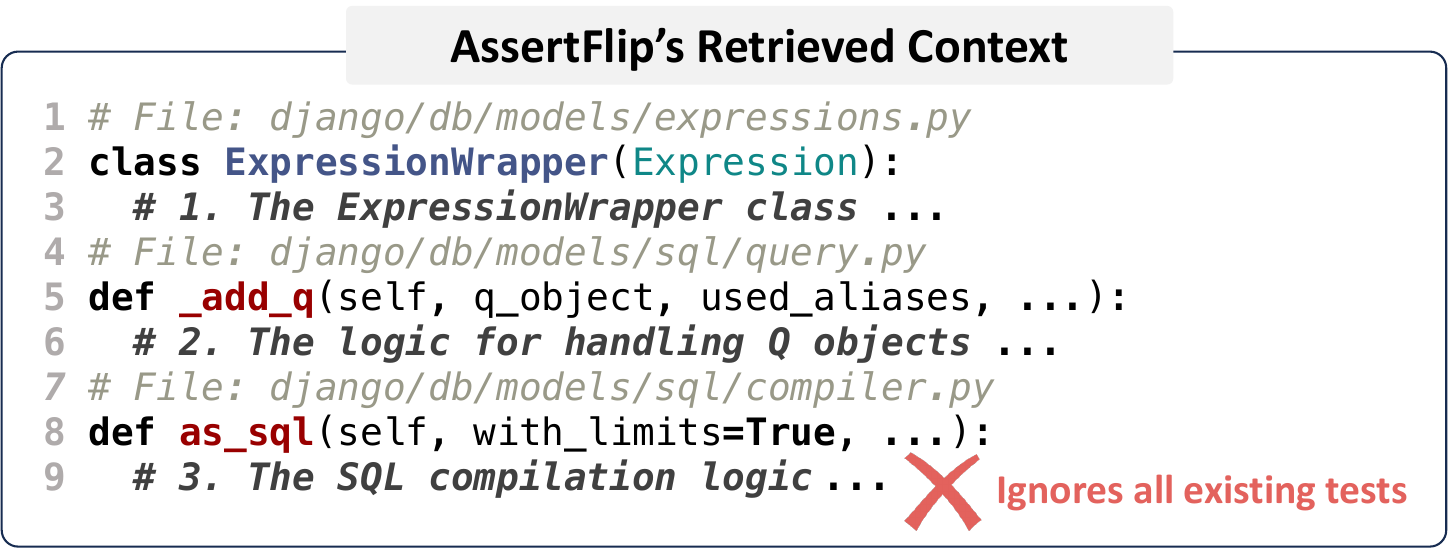}
    \caption{AssertFlip's Retrieval Failure Due to the Absence of Differentiation Between Code and Test.}
    \label{fig:assertflip_example}
  \end{subfigure}
  \hfill
  \begin{subfigure}{0.48\linewidth}
    \includegraphics[width=0.95\linewidth]{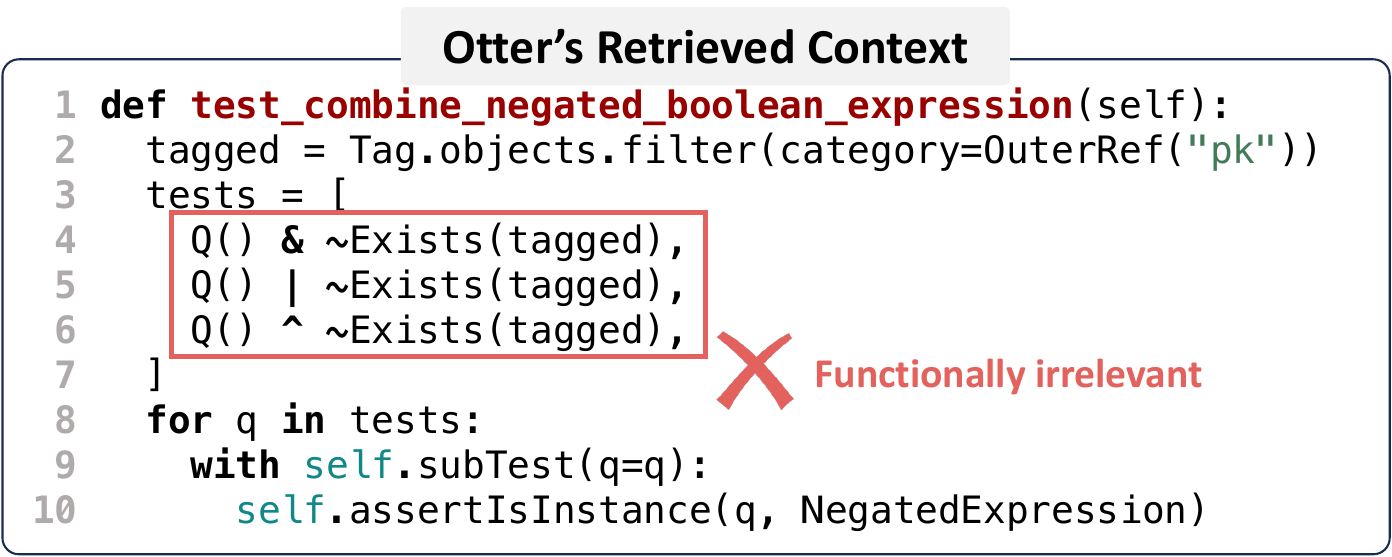}
    \caption{Otter's Retrieval Failure Due to the Neglect of Function Call Relationships.}
    \label{fig:otter_example}
  \end{subfigure}

  \begin{subfigure}{\linewidth}
    \includegraphics[width=\linewidth]{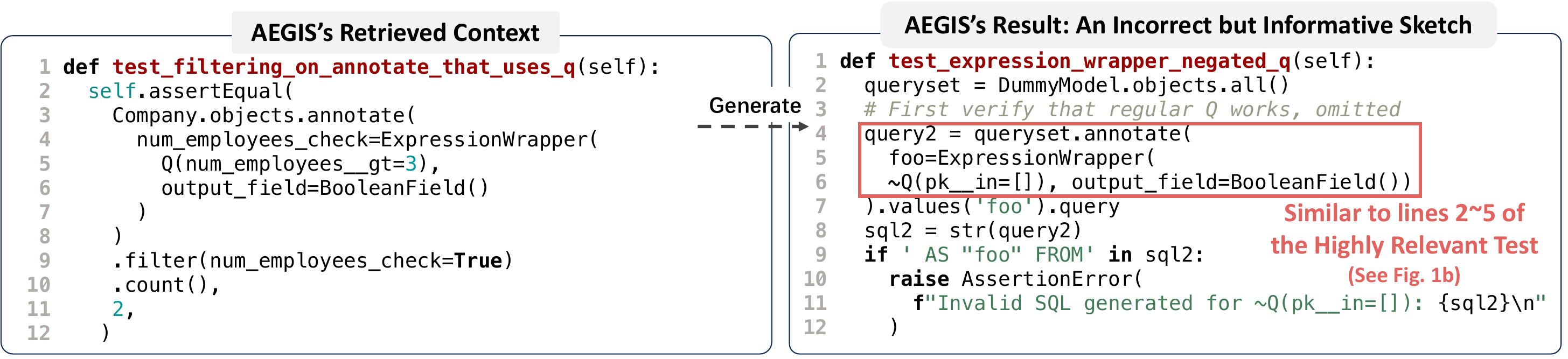}
    \caption{AEGIS's Retrieval Failure Due to the Lack of a Feedback from Generation.}
    \label{fig:aegis_example}
  \end{subfigure}
  \caption{Motivating Example (Continued from Figure~\ref{fig:motivating_example})}
  \label{fig:motivating_example_cont}
\end{figure}

To address the above limitations, we propose a novel \textbf{i}terative \textbf{Co}rrelation-aware \textbf{Re}trieval approach, \appname, for automated BRT generation.
For the limitation of the unified retrieval strategy, \appname adopts a two-stage process that handles production code and test separately. 
For production code retrieval, it queries a project-level code index based on keywords in the issue description and heuristic rules.
For test code retrieval, we introduce an iterative refinement loop.
Within each iteration, we apply a generation module to produce a sketch BRT, which serves as feedback to the retrieval.
The sketch is then compared against the existing test suite based not only on textual similarity but also on its function call patterns, which addresses the second and third limitations.
\appname continuously gains feedback from the generation phase to refine its context, progressively focusing on the most relevant tests.

To evaluate the effectiveness of \appname, we connect it to a BRT generator, 
which utilizes the high-quality context from \appname to prompt an LLM for generating BRTs. 
Our retriever plus this module achieved Fail-to-Pass rates of 42.0\% and 52.8\% using GPT-4o on SWT-bench Lite and TDD-bench Verified, respectively, representing relative improvements of 19.7\%--31.7\% over the best prior retrieval methods.
Notably, \appname is generator-agnostic, meaning it can be readily plugged into various generation modules to enhance the overall performance of BRT generation.
Furthermore, an ablation study demonstrated that every component of our framework is indispensable.

Our contributions can be summarized as follows:
\begin{itemize}
    \item We propose a novel, two-stage iterative context retrieval method for bug reproduction test generation. 
    Specifically, our method:
    1) Employs a specialized, two-stage approach for production code and test code;
    2) moves beyond simple textual similarity to capture function call relationships for deeper structural relevance;
    and 3) introduces a feedback loop from generation to retrieval using generated sketch BRTs, enabling iterative context refinement.
    \item We apply our retrieval method to enhance LLM-based test generation for bug reproduction. 
    We show that providing iteratively refined, high-quality context significantly improves the accuracy and validity of the generated tests.
    \item We demonstrate the effectiveness of our approach through extensive experiments. 
    On SWT-bench Lite and TDD-bench Verified, our framework achieves Fail-to-Pass rates of 42.0\% and 52.8\% respectively, validating the superiority of our proposed method.
\end{itemize}

\section{Motivation}

In this section, we illustrate the limitations of existing retrieval approaches through a real-world bug, ``django\_\_django-15213'', from the SWT-bench benchmark\cite{mundler2024swtbench}.
As shown in Figure \ref{fig:prob_decs}, this bug occurs when the system processes a negated empty condition object \texttt{~Q(pk\_\_in=[])}
, which produces invalid SQL and causes a crash.
The goal is to generate a test that reproduces this specific error automatically.
For this example, a straightforward solution exists. 
The solution requires retrieving a highly relevant test already present in the project: ``test\_empty\_expression\_annotation'', which provides a near-perfect template.
The correct test can then be produced by making a simple modification: adding a tilde symbol before the \texttt{Q} object on Line 4.
However, most existing methods cannot generate a test that successfully triggers this error. 
After our analysis, the root causes of failures
can be attributed to their retrieval, which supplies insufficient
context.

The first limitation of existing retrieval methods is their failure to differentiate between the distinct needs of retrieving production code and test.
For example, 
AssertFlip's \cite{khatib2025assertflipreproducingbugsinversion} retrieval, which uses a unified retrieval strategy, 
successfully retrieves several relevant code snippets but completely overlooks the distinct retrieval requirements for test cases, as shown in Figure \ref{fig:assertflip_example}.
It overlooks the need for test examples that demonstrate how to construct the scenario that triggers the bug.
Consequently, the generator receives insufficient context and produces an incorrect BRT.
In contrast, an approach that distinguishes between production code and tests can retrieve not only the faulty code but also the essential test templates required for bug reproduction.

The second limitation of existing retrieval methods is their reliance on textual semantics while ignoring the function call relationships.
For example, Otter's \cite{otter_ahmed_et_al_2025} retrieval employs a two-step workflow, which instructs an LLM to first identify relevant files,
then select the most relevant function names from that subset.
As shown in Figure \ref{fig:otter_example}, Otter's retrieval ultimately selects ``test\_combine\_negated\_boolean\_expression'' as the relevant test. 
This choice appears reasonable, as its name contains semantically aligned concepts like ``negated'' and ``expression''.
However, this reliance on ambiguous semantics
completely ignores the underlying function call relationships. 
The logic of the retrieved test centers on the interaction between \texttt{Q} and \texttt{Exists}, but never invokes the very components required to trigger the bug like \texttt{ExpressionWrapper}.
Consequently, the selected test is functionally irrelevant and serves only as a distraction for LLMs.
In contrast, an analysis of function calls confirms that ``test\_empty\_expression\_annotation'' is behaviorally more relevant, because it invokes the necessary methods for triggering the bug.
This highlights the necessity of analyzing function call relationships
to determine behavioral relevance.

The third limitation in existing methods is the absence of feedback from generation to retrieval.
For example, AEGIS \cite{aegis} incorporates an LLM-based search agent 
to perform context retrieval.
It located a closer but suboptimal test, ``test\_filtering\_on\_annotate\_that\_uses\_q'', as illustrated in Figure \ref{fig:aegis_example}, which still diverges from the essence of the bug.
Based on this suboptimal context, AEGIS generates a failed BRT.
However, this incorrect BRT can serve as a powerful new clue, as it successfully synthesizes the core bug pattern, providing valuable feedback from generation.
This feedback can then be used to perform a more targeted search, leading to the highly relevant test, ``test\_empty\_expression\_annotation'', that was initially missed.
This clearly demonstrates the critical importance of a generation-to-retrieval feedback iteration: by leveraging the sketch BRT from the generation phase as a new, particular query, the initial retrieval errors can be rectified and converge on the better context that static, one-way methods overlook.

The limitations mentioned above motivate us to propose \appname, a novel, correlation-aware approach.
Specifically, \appname is aware of three key correlations: 
 1) the complementary relationship between production code and test cases, which requires differentiated retrieval; 
 2) the interplay between textual semantics and function call structures, which must be jointly assessed for behavioral relevance; and 
 3) the powerful feedback correlation between the generation and retrieval phases, which enables iterative refinement.
By being aware of these correlations, \appname systematically alleviates the failures of prior approaches.

\section{Approach}

\begin{figure}[t]
    \centering
    \includegraphics[width=\linewidth]{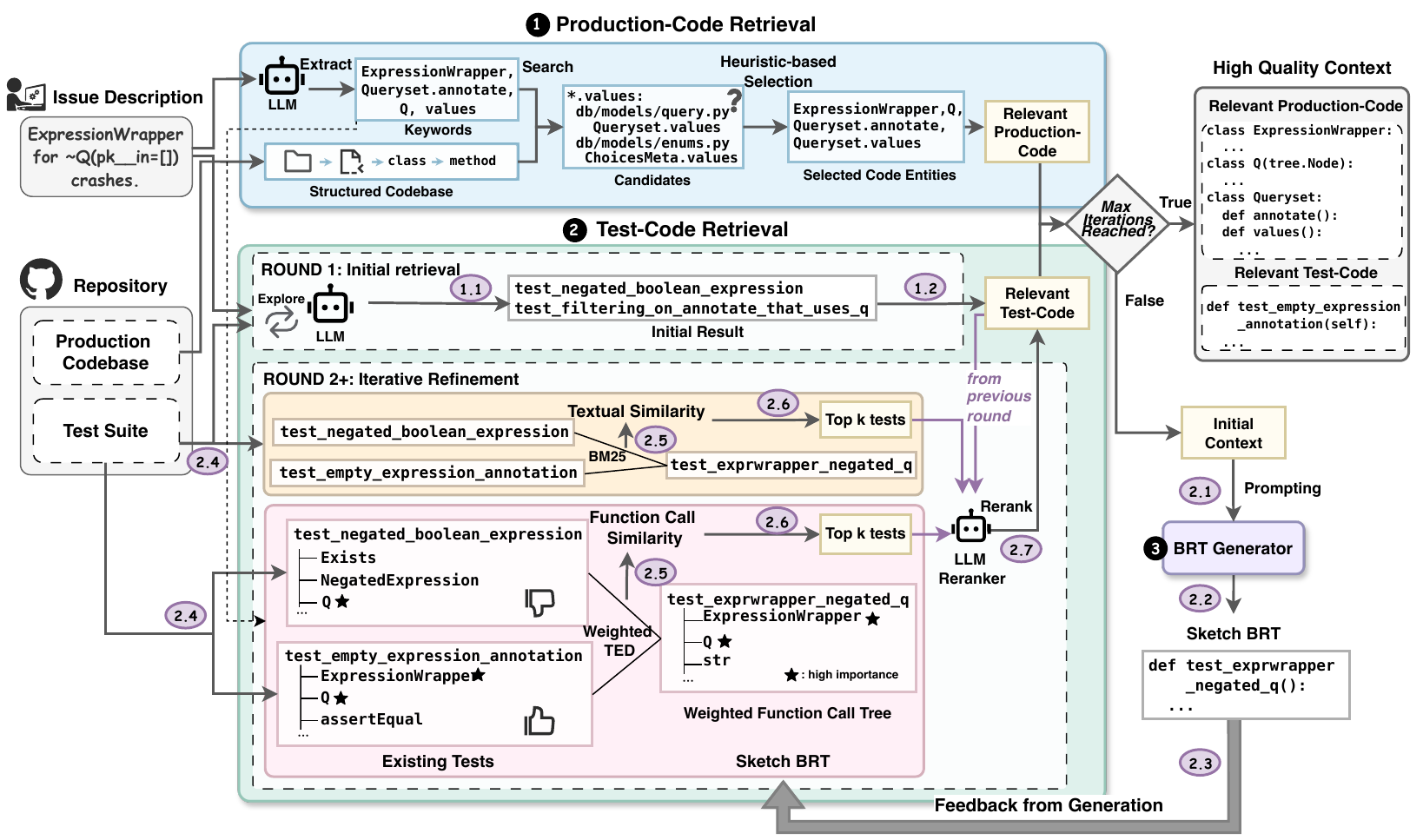}
    \caption{Overview of \appname.}
    \label{fig:framework}
    \vspace{-1em}
\end{figure}

In this section, we introduce the framework of \appname.
As illustrated in Figure \ref{fig:framework}, the overall process of \appname takes an issue description and a code repository as input to produce a high-quality context as output.
\appname 
consists of three core modules: 
\ding{202} a Production-Code Retrieval module, uses the issue description and the production codebase as input to retrieve relevant production code;
\ding{203} a Test-Code Retrieval module, which takes the issue description and the test suite to retrieve relevant tests;
and \ding{204} a BRT Generator, an LLM-based framework, generates a sketch BRT based on the issue description and the retrieved context. 

The pipeline of \appname is as follows:
The Production-Code Retrieval module first extracts keywords (i.e., code-related entities) from the issue description. 
These keywords are then searched against a hierarchical structure of the repository's files, classes, and methods to gather a set of candidates. 
Finally, a heuristic-based selection process pinpoints the most relevant code entities to form the relevant production code.
The Test-Code Retrieval module executes an iterative process.
In the initial round, it utilizes an LLM to explore the test suite based on the issue description and retrieve an initial set of relevant tests.
From the second round onwards, the module begins an iterative refinement loop. 
In each iteration, it leverages the retrieved production code and the test code from the previous round to prompt the BRT Generator for a sketch BRT.
This sketch then serves as input to the Test-Code Retrieval module. 
By calculating textual and function-call similarities between the sketch BRT and existing tests, the module selects a top-k test list for each metric. 
These tests are then fed into an LLM reranker, which selects and outputs the most relevant tests.
This iterative process terminates once the maximum number of iterations is reached. The retrieved code and tests are combined as the context retrieved by \appname.

\subsection{Production-Code Retrieval}
\label{sec:code-retrieval}
The production-code retriever takes as input an issue description and the project's production codebase, and outputs a set of relevant production code 
The process involves three main steps: keyword extraction, keyword
search, and heuristic-based selection.

\begin{table}[t]
    \centering
    \caption{Ambiguity in keyword retrieval: Retrieval results for the keywords ``annotate'' and ``values''.}
    \label{tab:keyword-example}
    \resizebox{0.7\linewidth}{!}{%
    \begin{tabular}{lll}
        \toprule
        \textbf{keywords} & \textbf{file} & \textbf{parent} \\
        \midrule
        annotate & django/db/models/query.py & Queryset \\
        \hline
        \multirow{3}{*}{values} & django/db/models/query.py  & Queryset \\
          & django/db/models/enums.py  & ChoicesMeta    \\
         & django/contrib/sessions/backends/base.py & SessionBase \\
        \hline
    \end{tabular}
    }
\end{table}

\textbf{Keyword Extraction.} We utilize LLMs to extract potential code-related entities from the issue description, including function, class, and module names.
For example, as shown in Figure 3, we extract keywords such as \texttt{ExpressionWrapper}, \texttt{Q}, \texttt{annotate}, and \texttt{values}.
To precisely identify code elements from a issue description, we designed a custom prompt (provided in our appendix) that guides the model to identify code elements while filtering out irrelevant information.
The prompt instructs the model to infer the full entity path (e.g., \texttt{Queryset.annotate}).
Recognizing that this inference can be imprecise, if a search for the full path yields no results, we incorporate a fallback strategy that uses only the final entity name (e.g., \texttt{*.annotate}) for the subsequent search. 

\textbf{Keyword searching.} We parse the code repository into a hierarchical structure of files, classes, and methods, where each of these elements is treated as a searchable node. 
By performing string matching between the extracted keywords and the node names in this structural tree, we obtain an initial set of candidates. 
However, common keywords can appear in multiple locations, 
such as ``\texttt{values}'' extracted from the issue description shown in Figure \ref{fig:framework}. 
As detailed in Table \ref{tab:keyword-example}, this results in a high number of irrelevant candidates, which reduces retrieval accuracy.

\textbf{Heuristic-based Selection.}
To address this ambiguity, we designed a heuristic-based selection mechanism to identify the most relevant nodes, as detailed in our appendix.
This mechanism considers three key factors:
\textit{(1) Node Uniqueness}: 
Keywords that match only a few nodes are considered more specific and are given higher priority.
For instance, in Table \ref{tab:keyword-example}, ``Queryset.annotate'' matches a single node, making it a high-confidence result. 
\textit{(2) File Co-occurrence}:
A file is considered more relevant if it contains nodes matched by multiple different keywords.
In the example from Table~\ref{tab:keyword-example}, the first candidate for \texttt{values} is in the same file (query.py) as \texttt{annotate}, suggesting they are more relevant. 
\textit{(3) Structural Relationship}: 
Nodes that share a parent in the code hierarchy, such as belonging to the same class, are considered more relevant.
As seen in Table~\ref{tab:keyword-example}, the parent of the first \texttt{values} candidate is \texttt{Queryset}, same as \texttt{annotate}, indicating a strong structural relationship. 

Through this process, the Production-Code retrieval module efficiently filters a large codebase to identify the code snippets most relevant to the issue description, providing high-quality context for the subsequent test generation phase.

\subsection{Test-Code Retrieval}
\label{sec:test-retrieval}
The input for this module is the issue description and the project's test suite. 
In iterative rounds, it also takes the sketch BRT generated from the previous round. 
Its output is a set of highly relevant test cases. 
The retrieval process begins with an initial search and is followed by iterative refinement.

\subsubsection{Initial Retrieval}
\label{subsec:initial-retrieval}
In the first round, we employ an LLM for an exploratory search.
To facilitate this exploration, we provide the LLM with a set of tool functions: \textit{list\_root} and \textit{list\_folder} to list directory structures, \textit{list\_tests} to identify test functions within files, and \textit{read\_function} to retrieve the production code of specific functions. 
Guided by the issue description, the LLM identifies an initial set of tests. 
However, this initial LLM-guided search has inherent limitations. 
The model may terminate its exploration prematurely once it discovers a test that appears textually relevant and overlook other behaviorally relevant tests that are less textually relevant.
Consequently, the initial set of tests is often suboptimal and may not be truly relevant to the bug's root cause.

\subsubsection{Similarity Calculation for Iterative Refinement}
\label{subsec:similarity-calc}
To overcome the limitation of the initial search and refine the suboptimal tests, we introduce an iterative process driven by a comprehensive similarity calculation.
We initiate this process by first constructing an initial context that combines the initially retrieved tests with the relevant production code from Section~\ref{sec:code-retrieval}.
Based on this context, we generate a sketch BRT (e.g., ``test\_exprwrapper\_negated\_q'' as shown in Figure~\ref{fig:framework}).
Starting from the second round, this sketch BRT is then used as a query to find more relevant tests by calculating its similarity to all existing tests in the test suite.
We assess similarity across two dimensions: textual and function call similarity.

For textual similarity, we use the BM25 algorithm to compare the entire code of the generated BRT with existing tests, as well as comparing test function names. .
We treat test names as a distinct feature because they often concisely summarize a test's objective and contain keywords highly relevant to the issue description.
The final textual similarity score is a combination of the name similarity and the code similarity, calculated as $ \operatorname{Similarity_{textual}} = \text{sim}_\text{name} + \text{sim}_\text{code} $

However, relying solely on textual similarity can be misleading. 
For instance, in the example from Figure \ref{fig:framework}, the sketch BRT may be textually closer to ``test\_negated\_boolean\_expression''.
However, ``test\_empty\_expression\_annotation'' is the most relevant test case in this example.
We observe that the function call pattern of the most relevant test cases and the sketch BRTs are often more similar, which 
motivates our introduction of function call similarity to measure the behavioral resemblance between tests.
Specifically, we model these relationships for each test as a pruned Function Call Tree, rooted at the test function itself, and stop expanding the call path once it enters a production module. 
We chose a pruned tree over a complete call graph for two reasons. 
First, it allows us to focus on the test's logic while reducing noise by excluding deep, irrelevant implementation details from the application code. 
Second, it reduces computational cost, as comparing full project-level call graphs is prohibitive. 
Furthermore, since test cases are typically independent and do not call one another, this pruning process naturally results in a tree structure rather than a graph.

We employ a Weighted Tree Edit Distance (TED) \cite{tai1979tree} 
to quantify the similarity between two call trees. 
Unlike standard TED, which treats all nodes equally, we consider the different importance of each function.
For instance, a function explicitly mentioned in an issue description is far more significant than a common utility function.
Thus, we assign a weight W(f) to each node (i.e., function f) in the call tree, which serves as the cost for edit operations (e.g., insertion, deletion).

The weight W(f) for any given function f is determined by the following piecewise function:
\begin{equation}
  W(f) = \begin{cases} 1.0 & \text{if } f \in S_{\text{keyword}} \\ \alpha + (0.9-\alpha) \times \frac{\text{IDF}(f)}{\text{IDF}_{\max}} & \text{otherwise} \end{cases} 
 \label{eq:weight}
\end{equation}
This formula uses a hybrid strategy, with its components detailed as follows: 
\textit{(1) Keyword Functions}:
Since these function names are directly linked to the bug description, we consider them the strongest indicators of a test's intent, such as ``\texttt{ExpressionWrapper}'' and ``\texttt{Q}''. 
The keyword set, $S_{\text{keyword}}$, is directly derived from the code entity names extracted from the issue description by the Relevant Code Retriever in Section \ref{sec:code-retrieval}. 
Accordingly, any function $f$ belonging to this set is assigned a fixed maximum weight of $1.0$.
\textit{(2) Non-Keyword Functions}:
If it appears in many tests (e.g., ``assertEqual'', ``str''), it is less distinctive and should receive a lower weight. 
Conversely, a function that appears in only a few specific tests is more likely to represent the core logic of those tests and should be assigned a higher weight.
Therefore, we leverage Inverse Document Frequency (IDF) \cite{sparck1972statistical} to assess their importance. 
The resulting IDF score is then normalized to derive the final weight. 
The parameter $\alpha$ serves as a base weight to ensure that common functions also have a minimum contribution to the edit cost. 
The term $(0.9-\alpha)$ scales the IDF score, capping the weight of non-keyword functions at 0.9 to distinguish them from exact keyword matches (1.0).
In this work, we set the default $\alpha$ to 0.1. 
This configuration ensures that the variable IDF component remains dominant over the base weight and that similarity scores are driven by term specificity.

We apply the classic Zhang-Shasha algorithm \cite{doi:10.1137/0218082} for computing the weighted TED ($TED_w$), between the call tree of the generated test ($T_g$) and an existing test ($T_e$), then we normalize this distance into a similarity score using the following formula:
$$
\operatorname{Similarity_{func\_call}}(T_{g}, T_{e}) = 1 - \frac{TED_w(T_{g}, T_{e})}{W(T_{g}) + W(T_{e})}
$$
The denominator is the sum of the total weights of both trees, representing the theoretical maximum edit distance. 
A tree's total weight is the sum of all its node weights, $W(T)=\sum_{f \in T}{W(f)}$.
This transforms the distance into a similarity score, where a value closer to 1 indicates a higher degree of similarity in the call structure.

\subsubsection{LLM-Based Reranking}
\label{subsec:reranker}
In the final stage of each retrieval iteration, we use an LLM to synthesize the results and produce a final ranked list of relevant tests.
This phase is necessary to filter the large set of potentially noisy candidates from the initial retrieval and to ensure the final selection aligns with the original issue description, not just the sketch BRT.
LLM is suited for this task due to its ability to perform analysis across both structured test code and natural language issue descriptions.

The reranker takes three sets of test cases as input: the top-k tests identified by textual similarity, those from function call similarity, and the tests selected in the previous iteration.
By providing all three sets, we enable the LLM to act as an expert judge, considering not only the textual and behavioral signals from the current iteration but also the historical context from the previous one. 
This process allows the reranker to capture complex relationships that a simple score combination might miss, resulting in a more sufficient and helpful context for the next generation stage.
While LLMs may make biased judgments, we employ them in the loop for their automation and powerful reasoning capabilities.
As our evaluation in Section \ref{sec:RQ4} shows, such incorrect judgments are rare.

The process of generating a test sketch, using it to refine retrieving and reranking a new set of tests, concludes one iteration of test retrieval. 
Upon reaching the maximum number of iterations, the reranker's final output is taken as the definitive relevant test code. 
This test code, together with the relevant production code from Section \ref{sec:code-retrieval}, constitutes the final high-quality context.

\subsection{BRT Generator}
\label{subsec:generator}
This module is responsible for generating the BRT using the high-quality context provided by the retrieval modules. 
The main part of this module is a carefully designed prompt, the full version of which is provided in our appendix.
This prompt guides the entire generation process: it first instructs the model to analyze the issue description, explicitly identify the Observed Behavior (OB) and the Expected Behavior (EB). 
The model is then directed to synthesize this analysis with the provided contextual information. 
Based on this comprehensive understanding, the model generates the sketch BRT.

\section{Experimental Setup}
This section systematically evaluates the effectiveness of our proposed approach. We design experiments around three key research questions (RQs):
\begin{itemize}
    \item RQ1: How effective and efficient is \appname in supporting end-to-end bug reproduction compared with different retrieval methods?
    \item RQ2: How does \appname perform when paired with a basic generator, compared to state-of-the-art bug reproduction methods?
    \item RQ3: What is the contribution of each component of \appname to the overall performance?
    \item RQ4: How accurate is \appname at locating target tests compared to baseline retrieval methods and its ablation variants?
    \item RQ5: How do key hyperparameters impact the performance of \appname?
\end{itemize}

\subsection{Dataset}

We conduct experiments on two benchmarks, i.e., SWT-bench Lite~\cite{swtbench-lite} and TDD-bench Verified~\cite{ahmed2024tddbenchverifiedllmsgenerate}. 
Both of them are widely used in prior work~\cite{otter_ahmed_et_al_2025,ahmed2025executionfeedbackdriventestgeneration,aegis,khatib2025assertflipreproducingbugsinversion}.
SWT-bench Lite and TDD-bench Verified, respectively, include 276 and 449 real-world bug instances collected from 12 popular open-source projects. 
Each instance provides a detailed issue description, the buggy version of the repository, the fix patch, and a ground-truth BRT patch that fails on the buggy version and passes on the fixed version.

\subsection{Evaluation Metrics}
\subsubsection{Bug Reproduction Performance}
To evaluate bug reproduction performance in our analyses for RQ1, RQ2, RQ3, and RQ5, we adopt the primary metrics established by the SWT-bench benchmark, which are as follows:
\begin{itemize}
    \item \textbf{Fail-to-Pass Success Rate (F\textrightarrow P(@k))}: 
    It measures the ability to reproduce a bug successfully. 
    A reproduction is considered successful if a generated test fails on the buggy version, passes on the fixed version, and introduces no new failures on the fixed version (×\textrightarrow F). 
    F\textrightarrow P(@k) is the percentage of bug instances where at least one of the top k candidates is an F\textrightarrow P test.
    \item \textbf{Patch Coverage Score}: 
    This metric measures the extent to which a generated test covers the developer's fix patch.
    Given the specific implementations of the benchmarks, we adopt the coverage metric provided by each.
    For SWT-bench Lite, we report Delta Change Coverage ($\Delta \mathcal{C}$)~\cite{mundler2024swtbench}, calculated as the percentage of modified lines in the patch that are newly covered by the generated test.
    For TDD-bench Verified, we report Adequacy~\cite{ahmed2024tddbenchverifiedllmsgenerate}, which measures the ratio of added and deleted lines in the ground-truth patch that are covered by the test execution. 
    In our evaluation, we refer to both metrics collectively as Patch Coverage.
\end{itemize}

\subsubsection{Relevant Test Retrieval Performance}
To evaluate the test retrieval module's (detailed in Section ~\ref{sec:test-retrieval}) 
ability to locate target tests in our analysis for RQ4, we use three standard metrics from the code search domain:

\begin{itemize}
    \item \textbf{Mean Average Precision (MAP)}: This metric considers the rank of all relevant items, reflecting the overall quality of the ranking by measuring both precision and recall.
    \item \textbf{Mean Reciprocal Rank (MRR)}: This metric measures the rank of the first correctly retrieved item. A higher MRR indicates that a correct item was found earlier in the list.
    \item \textbf{Hit Rate @k (Hit@k)}: This metric measures the percentage of queries for which at least one correct item is found within the top k results.
\end{itemize}

\subsection{Comparative Methods}
\subsubsection{Retrieval Baselines}
To compare the capability of \appname against existing methods in retrieving context for bug reproduction in our analyses for RQ1 and RQ4, we selected the following diverse set of retrieval baselines:
\begin{itemize}
    \item \textbf{BM25} \cite{mundler2024swtbench}: A traditional retrieval method used in its ZeroShot and ZeroShotPlus configurations. We use the official results for this baseline as provided by the SWT-bench.
    \item \textbf{AEGIS's Retriever} \cite{aegis}: AEGIS is an agent-based BRT generation framework, which employs a search agent that uses tools to explore the repository. 
    While AEGIS itself generates BRT in script mode rather than the unit tests used by our generator and other methods(unit test mode), its retrieval component is suitable for comparison.
    We directly extract the retrieved context from the official execution logs of AEGIS on the SWT-bench leaderboard.
    \item \textbf{AssertFlip's Retriever} \cite{khatib2025assertflipreproducingbugsinversion}: Employs a hierarchical retrieval method focused exclusively on retrieving production code. It first locates relevant source files and then identifies specific functions within them to serve as context. We use their officially published retrieval results, which are available in their public GitHub repository.
    \item \textbf{Otter's Retriever} \cite{otter_ahmed_et_al_2025}: Also uses a hierarchical retrieval approach, but separates the process into two stages to retrieve relevant code and test contexts independently. Since its official retrieval results are not publicly available, we faithfully re-implemented this method based on the paper's description and prompts, using GPT-4o.
\end{itemize}

\subsubsection{BRT Generation Baselines}
In our analysis for RQ2, we compare the end-to-end performance of \appname paired with a basic generator against several state-of-the-art BRT generation methods reported on the SWT-bench leaderboard, including:
\begin{itemize}
    \item \textbf{ZeroShotPlus}\cite{mundler2024swtbench}: Proposed by SWT-bench, combining issue descriptions with BM25-retrieved context to directly prompt LLMs for test generation, utilizing a structured diff format to ensure the output is syntactically valid.
    \item \textbf{LIBRO}\cite{libro_kang_yoon_yoo_2023}: Generates multiple candidate tests with few-shot prompting and selects the best one using execution feedback.
    \item \textbf{Amazon Q Developer Agent}\cite{amazonqweb}: A commercial, closed-source method with unpublished implementation details.
    \item \textbf{AssertFlip} \cite{khatib2025assertflipreproducingbugsinversion}: First generates tests that pass on buggy behavior, then flips assertions to ensure failures on the buggy version.
    \item \textbf{OpenHands}\cite{wang2025openhands}: An open-source AI agent, adapted for bug reproduction, that tackles complex software engineering tasks by creating and executing a detailed plan of action.
    \item \textbf{e-Otter++} \cite{ahmed2025executionfeedbackdriventestgeneration}: A feedback-driven method that repairs tests using execution feedback, diversifies candidates with prompting, and leverages generated patches to assist final selection.
\end{itemize}

\subsection{Implementation Details}
\label{subsec:implementation}
We utilized three representative LLMs: 
(1) GPT-4o (gpt-4o-2024-08-06),  which is a state-of-the-art closed-source model and is widely used by recent studies in this field,
(2) DeepSeek-V3, which is a large-scale open-source model, 
and (3) Qwen3-32B, which represents a medium-scale open-source model.
To ensure deterministic and focused outputs during these steps, the model's temperature was set to 0 for keyword extraction, initial test retrieval, sketch generation, and reranking.
In Test-Code Retrieval (Section \ref{sec:test-retrieval})
, the number of max iterations (mentioned in Section \ref{subsec:reranker})
is set to 3.
This means the model generates a sketch BRT and reranks the candidate tests three times in succession. 
In the reranking stage (Section \ref{subsec:reranker}), LLMs are instructed to select no more than five tests, enforced programmatically. The justification for this parameter choice is explained in RQ5.

To evaluate the end-to-end BRT generation performance(i.e., RQ1, RQ2, RQ3 and RQ5), we applied \appname to a basic generator, which is inspired by LIBRO~\cite{libro_kang_yoon_yoo_2023}, a LLM-based BRT generation framework.
Since LIBRO was initially implemented for Java, we adapted it for the Python-based SWT-bench benchmark, involving a four-stage pipeline: prompt-engineering, generation, post-processing, and selection\&ranking.
The prompt-engineering stage follows the methodology described in Section \ref{subsec:generator}, where we integrate the retrieved context into a structured prompt to guide the LLM's generation. 
Following prior work~\cite{libro_kang_yoon_yoo_2023, khatib2025assertflipreproducingbugsinversion}, we perform BRT generation for each sample by creating a new test function and inserting it into an existing test file, regardless of whether the bug requires modifying an existing test or adding a new one.
Next, the post-processing stage adapts LIBRO's original Java-based logic for our Python environment, which includes fixing import statements and ensuring test name uniqueness by renaming any conflicting ones.
These steps are crucial for producing valid and executable tests.
Finally, the selection\&ranking stage identifies the best candidate by applying LIBRO's heuristic-based selection approach, which involves clustering candidates by failure type and ranking them based on relevance to the issue description.
In the basic BRT generator, we set the model's temperature to 0.7 and generate 10 candidate tests, evaluating the performance on the top-k (k=1, 5) candidates, in line with prior work~\cite{libro_kang_yoon_yoo_2023}.

\section{Experimental Results}
\subsection{RQ1: Comparison with Retrieval Baselines}
\label{sec:RQ1}

To answer RQ1, we evaluate the effectiveness of \appname by comparing it against several baselines. 
We conduct experiments on the two benchmarks i.e., SWT-bench Lite and TDD-bench Verified, with the three models, i.e., GPT-4o, Deepseek-V3, and Qwen3-32B.
To isolate the impact of the retrieval component and ensure a fair comparison, each retrieval method is integrated with the same basic BRT generator detailed in \ref{subsec:implementation}, with the only variable being the retriever itself.
To evaluate both the effectiveness and efficiency of each retriever, we report the Fail-to-Pass rates for the top-1 and top-5 selected candidates, as well as the average context length in tokens and the corresponding total input cost based on experiments with GPT-4o.
The cost is calculated by summing the individual GPT-4o input costs for each instance across the entire dataset.

\begin{table}[t]
    \centering
    \caption{Bug reproduction performance and context efficiency comparison of different retrieval modules on SWT-bench Lite and TDD-bench Verified. All retrieval methods are paired with the same basic BRT generator. Context efficiency metrics (length and cost) are derived from the GPT-4o experiments.}
    \label{tab:rq1}
    
    \resizebox{\textwidth}{!}{%
        \begin{tabular}{llcccccccccc}
            \toprule
            \multirow{2}{*}{\textbf{Benchmark}} & 
            \multirow{2}{*}{\textbf{Retrieval Method}} & 
            \multicolumn{2}{c}{\textbf{GPT-4o}} & 
            \multicolumn{2}{c}{\textbf{DeepSeek-V3}} & 
            \multicolumn{2}{c}{\textbf{Qwen3-32B}} & 
            \multicolumn{2}{c}{\textbf{Average}} & 
            \multirow{2}{*}{\makecell{\textbf{Avg. Context} \\ \textbf{Length}}} & 
            \multirow{2}{*}{\makecell{\textbf{Total Input} \\ \textbf{Cost (\$)}}} \\
            
            \cmidrule(lr){3-4} \cmidrule(lr){5-6} \cmidrule(lr){7-8} \cmidrule(lr){9-10}
            
             & & F$\to$P(@1) & F$\to$P(@5) & F$\to$P(@1) & F$\to$P(@5) & F$\to$P(@1) & F$\to$P(@5) & F$\to$P(@1) & F$\to$P(@5) & & \\
            \midrule
            
            \multirow{5}{*}{\makecell{SWT-bench \\ Lite}} 
             & BM25                   & 26.8 & 31.2 & 28.3 & 32.2 & 31.5 & 35.5 & 28.9 & 33.0 & 23,562 & 16.26 \\
             & AssertFlip's Retriever* & 27.9 & 34.1 & 33.3 & 36.6 & 30.4 & 33.7 & 30.5 & 34.8 & 4,324  & 2.98 \\
             & AEGIS's Retriever* & 30.8 & 39.1 & 33.7 & 37.7 & 33.0 & 37.3 & 32.5 & 38.0 & 3,766  & 2.60 \\
             & Otter's Retriever      & 31.9 & 40.2 & 33.3 & 35.1 & 34.4 & 38.0 & 33.2 & 37.8 & 2,418  & 1.67 \\
             & \appname               & \textbf{42.0} & \textbf{46.7} & \textbf{40.2} & \textbf{46.4} & \textbf{41.7} & \textbf{46.7} & \textbf{41.3} & \textbf{46.6} & 2,603 & 1.80 \\
            \midrule
            
            \multirow{3}{*}{\makecell{TDD-bench \\Verified}} 
             & BM25                   & 37.4 & 41.8 & 40.1 & 44.1 & 37.0 & 40.8 & 38.2 & 42.2 & 20,882 & 23.44 \\
             & Otter's Retriever      & 44.1 & 50.3 & 44.3 & 49.2 & 43.9 & 49.2 & 44.1 & 49.6 & 2,823 & 3.17 \\
             & \appname               & \textbf{52.8} & \textbf{55.2} & \textbf{51.4} & \textbf{55.7} & \textbf{48.1} & \textbf{51.0} & \textbf{50.8} & \textbf{54.0} & 2,468 & 2.77 \\
            \bottomrule
        \end{tabular}%
    }
    \raggedright  
    \footnotesize  
    \textit{Note:} Since the code of AEGIS's retriever and AssertFlip's retriever are not open-source, we utilized their provided retrieval artifacts (GPT-4o) as fixed input contexts when evaluating their performance with DeepSeek-V3 and Qwen3-32B.
    \vspace{-0.5em}
\end{table}

The results of our comparison with baseline retrieval methods are presented in Table~\ref{tab:rq1}.
\appname achieves the highest performance across all benchmarks in both metrics.
It reaches average F→P(@1) rates of 41.3\% on SWT-bench Lite and 50.8\% on TDD-bench Verified, outperforming the strongest baseline, Otter’s retriever, by 24.4\% and 15.2\%, respectively.
The performance advantage is more pronounced when paired with GPT-4o, where \appname outperforms Otter’s retriever by 10.1 percentage points (31.7\%) and 8.7 percentage points (19.7\%) on SWE-bench Lite and TDD-bench Verified, respectively.
In addition, this superiority generalizes to different models: whether paired with DeepSeek-V3 or Qwen3-32B, \appname consistently maintains its lead over the baselines on both datasets.
These results underscore the effectiveness of \appname in identifying more relevant and useful contexts for the BRT generator.

On SWT-bench Lite, the average context length of \appname is 2,603 tokens, which is 89.0\%, 39.8\%, and 30.9\% shorter than those of BM25 (over 25K), AssertFlip's retriever (4,324), and AEGIS's retriever (3,766). 
While \appname's context is 6.8\% longer than Otter's on SWT-bench Lite, it delivers vastly superior performance for a marginal increase in cost. 
Furthermore, on TDD-bench Verified, \appname achieves the highest F$\to$P rate (52.8\%) with the most concise context (2,468 tokens), which is 88.2\% shorter than BM25 (over 20k) and 12.6\% shorter than Otter's retriever (2,823).
This result underscores our method's ability to strike a better balance between context conciseness and completeness, achieving state-of-the-art performance in a highly cost-effective manner.

\rqbox{
\textbf{Answer to RQ1:} 
Experimental results show that \appname is more effective and efficient than existing retrieval methods for automated BRT generation.
The context provided by \appname enables the generation of higher-quality BRTs, achieving a relative improvement of 19.7\%--31.7\% with GPT-4o over the baseline in F\textrightarrow P(@1) across two benchmarks.
This performance advantage generalizes across both closed-source and open-source models.
Moreover, \appname offers more concise context, averaging $\sim$2,500 tokens, which is 12.6--89.0\% shorter than other methods.
}\label{finding1}

\subsection{RQ2: Comparison with BRT Baselines}
\label{sec:RQ2}

This research question aims to investigate if an advanced context retriever, when combined with a basic generator, can achieve an overall performance that matches or even surpasses that of more structurally complex, state-of-the-art bug reproduction systems.
Specifically, we pair \appname with the basic generator detailed in \ref{subsec:implementation}, without any additional feedback or iterative refinements, and evaluate this combination as a self-contained system. 
We then directly compare its performance with that of several state-of-the-art methods on the SWT-bench Lite leaderboard and TDD-bench Verified.
Furthermore, to evaluate the generalizability of \appname across different models, we selected not only GPT-4o but also DeepSeek-V3 and Qwen3-32B, as they are representative of the closed-source, open-source, and medium-scale models, respectively.

For a fair and transparent comparison, we adhere to the following data sourcing principles. 
For SWT-bench Lite, the F\textrightarrow P and Patch Coverage Score($\Delta C$) metrics for most baselines are taken directly from the official SWT-bench leaderboard. 
To ensure a fair comparison, we specifically use the result of e-Otter++ with GPT-4o as reported in e-Otter++'s paper~\cite{ahmed2025executionfeedbackdriventestgeneration}
, since the e-Otter++ entry on the SWT-bench leaderboard~\cite{swtbench-lite} uses a different model (i.e., Claude 3.7-sonnet). 
The $\Delta C$ metric for e-Otter++ with GPT-4o was not provided in its paper.
For TDD-bench Verified, the baseline results for e-Otter++ are sourced directly from its original paper~\cite{ahmed2025executionfeedbackdriventestgeneration}.
The cost for the baseline approaches was sourced from their respective papers.
Specifically, these data for OpenHands were calculated from its publicly available logs on the SWT-bench leaderboard. 
In contrast, these data for the Amazon Q Developer Agent remain unknown due to its closed-source nature.
The cost for DeepSeek-V3 and Qwen3-32B is not specified as they are open-source models.

\begin{table}[tb]
    \centering
    \caption{
        Comparison of \appname with the Basic Generator against state-of-the-art methods on the SWT-bench Lite and TDD-bench Verified datasets. 
        The Fail-to-Pass rate is measured at k=1 (F$\to$P(@1)). 
        Both F $\to$ P and Patch Coverage are reported as percentages (\%). `--' means unknown and `N/A' means Not Applicable.
    }
    \label{tab:rq2}
    
    \resizebox{0.8\linewidth}{!}{%
        \begin{tabular}{lllrrr}
            \toprule
            \textbf{Benchmark} & \textbf{Approach} & \textbf{Model} & \textbf{F $\to$ P} & \textbf{Patch Cov.} & \makecell{\textbf{Cost per}\\ \textbf{Instance (\$)}} \\ 
            \midrule
            
            \multirow{9}{*}{SWT-bench Lite} 
             & ZeroShotPlus            & GPT-4             & 9.4           & 21.5          & 0.29 \\
             & LIBRO                   & GPT-4             & 14.1          & 23.8          & 1.52 \\
             & OpenHands               & Claude 3.5 Sonnet & 28.3          & 52.4          & 0.35 \\
             & AssertFlip              & GPT-4o            & 35.1          & 44.2          & 0.96 \\
             & Amazon Q                & --                & 37.7          & \textbf{52.7} & --   \\
             & e-Otter++               & GPT-4o            & 40.2          & --            & 1.80 \\ 
             \cmidrule(l){2-6} %
             & \multirow{3}{*}{\makecell[l]{\appname +\\Basic Generator}} 
                                       & Qwen3-32B         & 41.7          & 52.6          & N/A  \\
             &                         & DeepSeek-V3       & 40.2          & 51.7          & N/A  \\
             &                         & GPT-4o            & \textbf{42.0} & 52.4          & \textbf{0.28} \\
            \midrule
            
            \multirow{4}{*}{TDD-bench Verified} 
             & e-Otter++               & GPT-4o            & 51.4            & 70.3            & 1.80   \\ 
             \cmidrule(l){2-6}
             & \multirow{3}{*}{\makecell[l]{\appname +\\ Basic Generator}} 
                                       & Qwen3-32B         & 48.1          & 73.0          & N/A  \\
             &                         & DeepSeek-V3       & 51.4          & 75.4          & N/A  \\
             &                         & GPT-4o            & \textbf{52.8} & \textbf{76.2} & \textbf{0.26} \\
            \bottomrule
        \end{tabular}%
    }
    \vspace{-0.5em}
\end{table}

The results in Table~\ref{tab:rq2} demonstrate the superior performance and efficiency of \appname.
\appname with GPT-4o achieves 42.0\% and 52.8\% in the core metric F\textrightarrow P on SWT-bench Lite and TDD-bench Verified, respectively. It surpasses all existing approaches, including the SOTA method e-Otter++ with GPT-4o (40.2\% and 51.4\%). 
In terms of Patch Coverage on SWT-bench Lite($\Delta C$), our approach achieves 52.4\%, which is on par with the record held by the commercial, closed-source tool Amazon Q (52.7\%). 
On TDD-bench, our approach achieves a high Patch Coverage of 76.2\%, outperforming e-Otter++ (70.3\%) by 5.9 percentage points.
Collectively, these results demonstrate that our retrieval method, even when paired with a basic generator, can outperform more complex systems on the bug reproduction task.

The average cost per instance of \appname + basic generator is merely \$0.26--\$0.28, far below the state-of-the-art approaches. 
For instance, when using the GPT-4o model, our approach is 6.4 and 3.4 times cheaper than e-Otter++ (\$1.80) and AssertFlip (\$0.96), respectively.
This indicates that by providing high-quality context, \appname effectively reduces the need for multiple calls to expensive LLMs, achieving high performance through a basic prompt-then-generate process.

Notably, our system (\appname + Basic Generator) achieves similar performance using LLMs of different sizes, i.e., success rate 41.7\% and 48.1\% (Qwen3, 32B) \textit{vs.} 40.2\% and 51.4\% (DeepSeek-V3, 671B) on SWT-bench Lite and TDD-bench Verified, respectively.
It demonstrates that the powerful capabilities of \appname are not limited to leading closed-source models.

\rqbox{
\textbf{Answer to RQ2}: 
Even when paired with a basic generator, the end-to-end performance of \appname significantly surpasses current state-of-the-art BRT generation approaches across both SWT-bench Lite and TDD-bench Verified, such as e-Otter++, with GPT-4o.
\appname offers better cost-effectiveness and is 6.4x cheaper than e-Otter++ when both use the same GPT-4o model. 
Furthermore, it demonstrates generalizability across different LLMs.
}\label{finding2}

\subsection{RQ3: Ablation Study}
\label{sec:RQ3}

To measure the contribution of each key component of \appname to the BRT generation performance, we conduct a comprehensive ablation study. 
To achieve this, we create four degraded variants of \appname by disabling individual components: \ding{172} \textbf{w/o Test}: The Test-Code retrieval module is removed. 
The context consists only of the retrieved production code. \ding{173} \textbf{w/o Code}: The Production-Code Retrieval module is removed. The context consists only of the retrieved test code. \ding{174} \textbf{w/o Iteration}: The iterative refinement loop is disabled. The context includes both the production code and the results from the initial test retrieval. \ding{175} \textbf{w/o Function-Call Similarity}: The function call analysis is excluded from the test re-ranking phase. 
To ensure fairness, the input for the re-ranking process is set to the top 10 textually similar tests and the selected tests
We then evaluate the end-to-end BRT generation of these four variants and compare them against the full \appname system.

\begin{table}[t]
  \caption{Contribution of each component to the overall bug reproduction performance on SWT-bench Lite and TDD-bench Verified. The Fail-to-Pass rate (F\textrightarrow P(@k)) is reported as a percentage (\%).}
  \vspace{-0.3em}
  \label{tab:rq3}
  
  \resizebox{0.8\textwidth}{!}{
  \begin{tabular}{l|cc|cc}
    \toprule
    \multirow{2}{*}{\textbf{Retrieval Component}} & \multicolumn{2}{c|}{\textbf{SWT-bench Lite}} & \multicolumn{2}{c}{\textbf{TDD-bench Verified}} \\
    \cmidrule(lr){2-3} \cmidrule(lr){4-5}
     & \textbf{F\textrightarrow P (@1)} & \textbf{F\textrightarrow P (@5)} & \textbf{F\textrightarrow P (@1)} & \textbf{F\textrightarrow P (@5)} \\ 
    \midrule
    
    \appname & \textbf{42.0} & \textbf{46.7} & \textbf{52.8} & \textbf{55.2} \\ 
    \midrule
    
    w/o Test 
    & 33.7 ($\downarrow$ 19.8\%) & 40.6 ($\downarrow$ 13.1\%) 
    & 35.2 ($\downarrow$ 33.3\%) & 39.9 ($\downarrow$ 27.7\%) \\
    
    w/o Code 
    & 36.6 ($\downarrow$ 12.9\%) & 41.7 ($\downarrow$ 10.7\%) 
    & 40.8 ($\downarrow$ 22.7\%) & 46.8 ($\downarrow$ 15.2\%) \\
    
    w/o Iteration 
    & 37.0 ($\downarrow$ 11.9\%) & 41.7 ($\downarrow$ 10.7\%) 
    & 44.5 ($\downarrow$ 15.7\%) & 49.2 ($\downarrow$ 10.9\%) \\
    
    w/o Function-Call Similarity 
    & 39.5 ($\downarrow$ 6.0\%) & 44.6 ($\downarrow$ 4.5\%) 
    & 46.5 ($\downarrow$ 11.9\%) & 52.1 ($\downarrow$ 5.6\%) \\
    
    \bottomrule
  \end{tabular}
  }
  \vspace{-0.5em}
\end{table}

The results of our ablation study, presented in Table~\ref{tab:rq3}, confirm that all of our designed components make positive and significant contributions to the overall performance of \appname.
Removing test case retrieval (w/o Test) leads to the most severe performance degradation across both benchmarks.
On SWT-bench Lite, F\textrightarrow P(@1) and F\textrightarrow P(@5) rates drop by 19.8\% and 13.1\%, respectively, while on TDD-bench, the degradation is amplified, reaching 33.3\% and 27.7\%. 
Removing code retrieval (w/o Code) also results in a significant drop of F\textrightarrow P(@1/@5) by 12.9\% and 10.7\% on SWT-bench Lite, and by 22.7\% and 15.2\% on TDD-bench Verified.
These results indicate that both test cases and production code play a critical role in BRT generation, but the former provides more critical guidance. 
This may be because relevant tests not only offer structural templates and assertion logic but also provide function call examples and object instantiation patterns, thereby partially encompassing the functional information found in the production code snippets.

When the iterative process is disabled (w/o Iteration), the F\textrightarrow P(@1) performance drops by 11.9\% on SWT-bench and 15.7\% on TDD-bench. 
It strongly shows the effectiveness of the feedback loop from a sketch BRT to the retrieval focus, as it enables the system to progressively mine for relevant information that may be missed in a single pass.

Removing the function-call similarity component (w/o Function-Call Similarity) results in a 6.0\% and 11.9\% F\textrightarrow P(@1) performance drop on two benchmarks. 
It validates our hypothesis that capturing behavioral relevance through function call structures, rather than relying solely on textual matching, enables a more accurate assessment of relevant test cases, thereby further enhancing context quality.

\rqbox{
\textbf{Answer to RQ3}: 
The ablation study confirms the effectiveness of all designs in \appname. 
Both relevant test cases and code snippets play a critical role in the final performance, and test cases make a more significant contribution than production code. 
The iterative refinement mechanism and the incorporation of function-call similarity also substantially contribute to the overall performance improvement.
}

\subsection{RQ4: Target Test Retrieval Accuracy Evaluation}
\label{sec:RQ4}

To evaluate the effectiveness of \appname in retrieving target tests, we built a dedicated dataset based on SWT-bench Lite and TDD-bench Verified. 
We define ``target tests'' as the test functions explicitly modified by developers during the bug-fixing process. 
These tests are considered the most directly relevant to the bug, as they are adapted to reproduce its behavior. 
Accordingly, we extracted these target tests from both datasets to serve as the ground truth for evaluating retrieval accuracy.

We compared the retrieval accuracy of \appname against several retrieval baselines that include test retrieval, including BM25, AEGIS's retriever, and Otter's retriever.
The former two are line-level retrievers that identify relevant code snippets, whereas the latter uses functions as its unit of retrieval.
Additionally, two ablation versions of \appname are included in the comparison, i.e., w/o Iteration and w/o Function-Call Similarity.

During the evaluation process, we defined two relevance criteria: 
\ding{172} file-level match, where the retrieved test is in the correct file; 
and \ding{173} function-level match, where either the retrieved test is the ground-truth test itself, or the retrieved line numbers fall within that function's range.
Based on these criteria, we calculated the MAP, MRR, and Hit@k (k=1, 5) for both file-level and function-level performance, providing a comprehensive measure of retrieval effectiveness.

\begin{table}[tb]
\caption{Comparison of retrieval accuracy for identifying target tests on SWT-bench Lite and TDD-bench Verified. Performance is measured at both file and function levels using MAP, MRR, and Hit@k. `--' indicate that the retrieval artifacts for the corresponding method on the specific benchmark were not available.}
\vspace{-0.2cm}
\resizebox{\linewidth}{!}{
\begin{tabular}{l|l|rrrr|rrrr}
\toprule
\multirow{2}{*}{\textbf{Match Type}} & \multirow{2}{*}{\textbf{Method}} & \multicolumn{4}{c|}{\textbf{SWT-bench Lite}} & \multicolumn{4}{c}{\textbf{TDD-bench Verified}} \\
 & & \textbf{MAP} & \textbf{MRR} & \textbf{Hit@1} & \textbf{Hit@5} & \textbf{MAP} & \textbf{MRR} & \textbf{Hit@1} & \textbf{Hit@5} \\
\midrule
\multirow{6}{*}{File Match} 
    & BM25                  & 0.198 & 0.210 & 0.177 & 0.250 & 0.139 & 0.148 & 0.115 & 0.198 \\ 
    & AEGIS's Retriever     & 0.524 & 0.552 & 0.531 & 0.572 & -- & -- & -- & -- \\ 
    & Otter's Retriever     & 0.591 & 0.621 & 0.596 & 0.676 & 0.618 & 0.655 & 0.620 & 0.723 \\ 
    & \appname              & \textbf{0.699} & \textbf{0.732} & \textbf{0.684} & \textbf{0.794} & \textbf{0.747} & \textbf{0.786} & \textbf{0.742} & \textbf{0.845} \\ 
    \cline{2-10}
    & w/o Iteration         & 0.561 & 0.576 & 0.559 & 0.596 & 0.569 & 0.608 & 0.592 & 0.629 \\
    & w/o Function Call Similarity  & 0.685 & 0.714 & 0.676 & 0.765 & 0.716 & 0.760 & 0.728 & 0.826 \\
\midrule
\multirow{6}{*}{Function Match} 
    & BM25                  & 0.049 & 0.051 & 0.037 & 0.052 & 0.012 & 0.013 & 0.002 & 0.018 \\ 
    & AEGIS's Retriever     & 0.180 & 0.214 & 0.162 & 0.287 & -- & -- & -- & -- \\ 
    & Otter's Retriever     & 0.319 & 0.381 & 0.346 & 0.426 & 0.311 & 0.370 & 0.300 & 0.474 \\
    & \appname              & \textbf{0.466} & \textbf{0.527} & \textbf{0.471} & \textbf{0.618} & \textbf{0.438} & \textbf{0.507} & \textbf{0.432} & \textbf{0.606} \\
    \cline{2-10}
    & w/o Iteration         & 0.344 & 0.380 & 0.346 & 0.434 & 0.316 & 0.371 & 0.310 & 0.451 \\
    & w/o Function Call Similarity & 0.449 & 0.510 & 0.463 & 0.588 & 0.418 & 0.485 & 0.413 & 0.592 \\
\bottomrule
\end{tabular}
}
\vspace{-0.5em}
\label{tab:rq4}
\end{table}

Table \ref{tab:rq4} demonstrates the superior accuracy of \appname in retrieving target tests and validates the contributions of its core components. 
Across all metrics at both the file and function levels, \appname significantly surpasses that of all baseline methods and its ablated versions. 
At the file-match level, \appname achieves a MAP of 0.699 on SWT-bench and 0.747 on TDD-bench, outperforming the best-performing baseline, i.e., Otter's retriever, by 18.3\% and 20.9\%.
Similarly, at the function-match level, \appname outperforms the strongest baseline with relative improvements of 46.1\% and 36.1\% in MAP and Hit@1 on SWT-bench Lite, alongside 40.8\% and 44.0\% on TDD-bench Verified.

Removing the iteration process (w/o Iteration) causes a drastic drop in accuracy across both benchmarks.
The function-level MAP decreases by 26.2\% on SWT-bench Lite and 27.9\% on TDD-bench Verified.
This confirms that the refinement loop is fundamental to discovering relevant contexts. 
Removing the function-call similarity analysis (w/o Function-Call Similarity) decreases function-level MAP by 3.6\% and 4.6\% on the two benchmarks, respectively.
This indicates that function-call analysis also plays an important role in the accurate retrieval of target tests.

Furthermore, we analyzed whether the LLM reranker (in Section \ref{subsec:reranker}) could cause target tests to be mistakenly overlooked. 
By comparing the tests fed into the reranker with its final output, we found that only 4\% of target tests were filtered out by the LLM. 
This low error rate confirms that the reranker is effective at preserving relevant tests while improving the overall context quality.

\rqbox{
\textbf{Answer to RQ4}:
\appname is demonstrably more accurate at retrieving relevant tests than both baseline methods and its ablated variants. 
Its high precision is primarily driven by its iterative refinement process, which is fundamental to its success. 
This core mechanism is further enhanced by function-call similarity analysis, which provides the necessary fine-grained precision to pinpoint the exact target test function.
}

\subsection{RQ5: Impact of Hyperparameters}
\label{sec:RQ5}

\begin{wrapfigure}[10]{r}{0.45\textwidth}
    \vspace{-1em}
    \centering
    \includegraphics[width=0.4\textwidth]{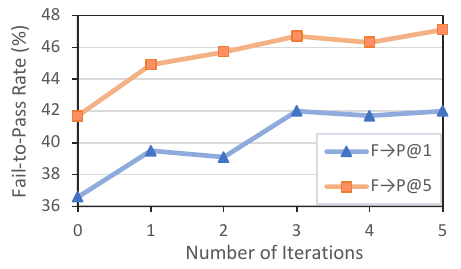}
    \vspace{-0.5em}
    \caption{Impact of Iteration Rounds.}
    \label{fig:iterations}
\end{wrapfigure}

We conducted a sensitivity analysis on three key hyperparameters using the SWT-bench Lite dataset: the number of retrieval iterations, the number of relevant tests provided as context, and the base weight used in function similarity calculation.

\noindent\textbf{Impact of Iteration Rounds.} We define an iteration as a cycle of generating a sketch BRT and using its feedback to refine the retrieval. 
Thus, ``0 iteration'' represents the initial, non-iterative retrieval result, while ``k iterations'' means k sketch BRTs were generated for feedback.
As shown in Figure \ref{fig:iterations}, the performance of \appname generally improves with more iterations.
Compared to the non-iterative baseline (0 iteration), iterative refinement yields a significant improvement in overall performance.
The F\textrightarrow P(@1) rate peaks at 42.0\% at the third iteration. 
Although the performance fluctuates slightly in subsequent rounds, with a minor improvement in the F\textrightarrow P(@5) rate, the overall performance effectively stabilizes around this peak.
This suggests that three feedback cycles are sufficient to gather the critical context for bug reproduction. 
Therefore, the default setting of three iterations is justified, balancing performance and efficiency.

\begin{wraptable}[7]{r}{0.45\textwidth}
    \vspace{-0.2em}
    \centering
    \caption{Impact of the Number of Provided Tests.}
    \label{tab:num_tests}
    \vspace{-0.5em}
    \resizebox{0.65\linewidth}{!}{
    \begin{tabular}{lrr}
    \toprule
    \textbf{\# Tests} & \textbf{F\textrightarrow P(@1)} & \textbf{F\textrightarrow P(@5)} \\ \hline
    1 & 34.4 & 41.7 \\
    \textbf{5} & \textbf{42.0} & 46.7 \\
    10 & 41.3 & \textbf{47.1} \\
    \bottomrule
    \end{tabular}
    }
\end{wraptable}

\noindent\textbf{Impact of the Number of Provided Tests.} We investigated the effect of the number of retrieved tests. 
This is controlled during the reranking stage by limiting the number of tests selected as the final context. 
We evaluated the performance when providing the top 1, 5, and 10 tests.
Table \ref{tab:num_tests} shows that providing five tests yields the best performance on F\textrightarrow P(@1). 
The performance is lowest when only 1 test is provided, because the critical context may not be consistently ranked as the top-1 candidate. 
When the number is increased to 10, the F\textrightarrow P(@1) rate declines by 0.7\%, while the F\textrightarrow P(@5) rate increases by 0.4\%, remaining essentially level with the performance when providing five tests. 
Consequently, using five tests strikes the optimal balance between providing comprehensive context and maintaining model focus and cost-effectiveness.
Therefore, this default setting is reasonable.

\begin{wraptable}[7]{r}{0.45\textwidth}
    \centering
    \vspace{-1em}
    \caption{Impact of the base weight parameter $\alpha$.}
    \vspace{-0.5em}
    \label{tab:sensitivity_weights}
    \small
    \resizebox{0.75\linewidth}{!}{
    \begin{tabular}{lcc}
    \toprule
    \textbf{Configuration} & \textbf{F$\to$P(@1)} & \textbf{F$\to$P(@5)} \\
    \midrule
    $\alpha=0.1$ (Default) & 42.0 & 46.7 \\
    $\alpha=0.2$ & 40.9 & 46.3 \\
    $\alpha=0.3$ & 41.3 & 46.7 \\
    $\alpha=0.4$ & 40.9 & 45.2 \\
    \bottomrule
    \end{tabular}
    }
\end{wraptable}

\noindent\textbf{Impact of Function Similarity Weights.}
We set the base weight $\alpha$ in Formula~\ref{eq:weight} (Section \ref{sec:test-retrieval}) to be 0.1 by default.
To analyze the model’s sensitivity to $\alpha$, we evaluate \appname with the basic generator by varying $\alpha$ from 0.1 to 0.4.
This upper bound was selected to ensure that the weight of the IDF component is larger than the base weight, so that the function weight is primarily driven by term specificity.
As detailed in Table \ref{tab:sensitivity_weights}, the F\textrightarrow P(@1) rate fluctuates minimally (Mean: 41.3\%, Std: 0.45\%) with different base weights, demonstrating the robustness of \appname to these configurations.
In addition, $\alpha=0.1$ yields optimal performance in our experiments. 
Therefore, we consider this default setting of $\alpha$ (0.1) to be reasonable.

\rqbox{
\textbf{Answer to RQ5: }
\appname's performance is influenced by the number of iterations and the number of provided tests. 
Conversely, the system demonstrates robustness regarding the function similarity weight ($\alpha$).
Our chosen default settings represent the optimal choice after considering both effectiveness and efficiency.
}

\section{Threats to Validity}
\noindent\textbf{Internal Validity.} The inherent non-determinism of LLMs poses a potential threat to the reproducibility of our results. 
To mitigate this threat, we intentionally set the temperature to 0 for all deterministic stages of the retrieval process, including keyword extraction, initial retrieval, sketch generation, and reranking. 
The second threat relates to the replication of baseline methods. 
For the Otter's retrieval baseline,
we had to rely on the methodology described in their original papers for replication, as their artifacts
were not publicly available. 
We have strictly followed the descriptions in their papers. 
The differences between our replication and their actual implementation, though not open-sourced, should be minor.

\noindent\textbf{External Validity.} First, our study shares a common threat with other contemporary work~\cite{aegis,otter_ahmed_et_al_2025,khatib2025assertflipreproducingbugsinversion,ahmed2025executionfeedbackdriventestgeneration}, i.e., the potential for data contamination~\cite{dataContamination2023,cao2024concerned} within the SWE-bench dataset. 
The pre-training corpora of the LLMs we used may have contained code or discussions related to the bugs in the benchmark. 
To assess the impact of this threat, we followed the methodology of Otter~\cite{otter_ahmed_et_al_2025} and conducted two quantitative analyses on the TDD-bench Verified dataset.
We analyzed the performance stability across issue creation years. 
We focused on the years 2017–2023, excluding earlier years with insufficient sample sizes (<5).
Performance across these issue creation years remains relatively stable (Mean: 57.16\%, SD: 11.75\%), suggesting that the model does not disproportionately favor older, potentially memorized issues.
We also measured the textual similarity between our generated tests and existing tests in the repository using the normalized Levenshtein similarity. 
We found that for 90\% of the instances, the similarity score remains below 0.56.
These findings confirm that the generator primarily relies on reasoning based on retrieved context rather than memorization, indicating that the impact of data contamination is limited.
Second, as our experiments were conducted exclusively on Python projects, our findings may not generalize to other languages.
However, Python is one of the most popular languages, and the architecture of \appname is language-agnostic and can be adopted to other languages.
Finally, our evaluation was performed solely on SWT-bench Lite, which may not fully encompass the diversity of bug types and project complexities encountered in real-world software development. 
However, this benchmark has been widely used in prior studies~\cite{aegis, otter_ahmed_et_al_2025}.
We intend to expand our evaluation to a broader range of programming languages, larger repositories, and more diverse bug scenarios in future work.

\section{Discussion}
\subsection{Comparison with e-Otter++}
Regarding e-Otter++ \cite{ahmed2025executionfeedbackdriventestgeneration}, its top performance on the SWT-bench Lite leaderboard (a 50.7\% success rate and 56.4\% change coverage) was achieved using Claude 3.7-sonnet. 
However, the cost per instance for e-Otter++ on Claude 3.7 is \$2.75, which is 9.8 times more than our method (\$0.28). 
Due to these significant cost constraints, we did not compare against it on Claude 3.7. 
Instead, we chose the widely used commercial model GPT-4o for a fair comparison. 
Our results show that when using the same GPT-4o model, the context retrieved by \appname, even when paired with a basic generator, surpasses the performance of e-Otter++ while incurring only 1/6.4 of the cost, highlighting the superior cost-effectiveness of our \appname.

\subsection{Failure Analysis}

To better understand the limitations of \appname, we conducted a qualitative analysis of failure cases of \appname paired with GPT-4o and the basic generator on SWT-bench Lite and TDD-bench Verified. 
We randomly sampled 50 failure instances from each benchmark (90\% confidence level, $\sim$10\% margin of error) and manually inspected them.
The failures were categorized into four primary types:

\begin{figure}
    \centering
    \includegraphics[width=0.85\linewidth]{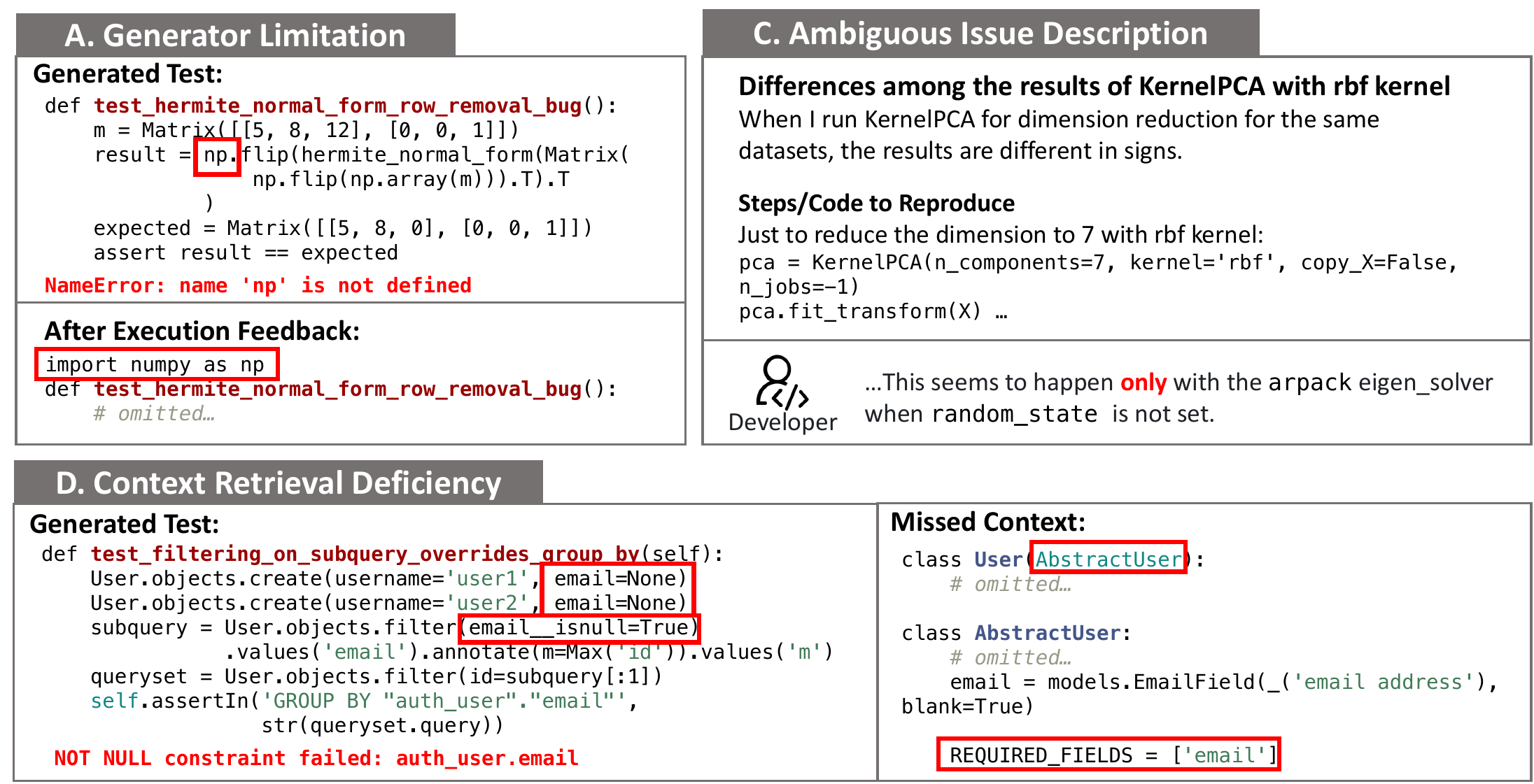}
    \caption{Examples of Failure Cases}
    \label{fig:failure_cases}
    \vspace{-1.3em}
\end{figure}

\noindent{\textbf{A. Limitation of Generator (SWT: 58\%, TDD: 84\%):}}

The majority of failures occurred when the retrieved context was sufficient, but the generator failed to output a correct BRT. 
These failures stem from the limited reasoning capabilities of the model or the simple one-pass generation process. 
For example, for \texttt{sympy\_\_sympy-23413} ( Figure~\ref{fig:failure_cases}A), the generated test failed to reproduce the issue simply due to a missing \texttt{numpy} import. 
With an execution feedback loop, the model could fix the \texttt{NameError} and successfully reproduce the bug.
This underscores the importance of execution feedback in BRT generation.
Future work could employ agentic workflows with self-correction capabilities to improve BRT generation.

\noindent{\textbf{B. Selection Failure (SWT: 18\%, TDD: 10\%):}} 

In these cases, the generator produced a correct BRT within the candidate list, but the selection module failed to rank it as the top-1 choice. 
This suggests that future work should focus on improving the ranking heuristics to better identify the correct test among candidates.

\noindent{\textbf{C. Ambiguous Issue Description (SWT: 14\%, TDD: 0\%):}}

Some failures stem from the quality of the issue description itself rather than the generator.
For instance, for \texttt{scikit-learn\_\_} \texttt{scikit-learn-13241}(Figure~\ref{fig:failure_cases}C), the description failed to mention that the bug only occurs when \texttt{eigen\_} \texttt{solver=`arpack'}.
Notably, no sample from TDD-bench Verified belongs to this category, as the issue descriptions in TDD-bench Verified are manually verified and filtered for clarity.

\noindent{\textbf{D. Context Deficiency (SWT: 10\%, TDD: 6\%):}}

In a minority of cases, the retrieval module failed to locate the critical context. 
For example, for \texttt{django\_\_django-11797}(Figure~\ref{fig:failure_cases}D), the retrieval failed to fetch the parent class \texttt{AbstractUser}.
Lacking this information, the generator failed to recognize that the \texttt{email} field has a \texttt{NOT NULL} constraint, causing an \texttt{IntegrityError} during the test setup. 
Future work could consider capturing inheritance hierarchies for context retrieval.

\section{Related Work}

\textbf{Benchmarks for Bug Reproduction.}
The systematic evaluation of LLM-based bug reproduction relies on specialized benchmarks.
SWT-bench\cite{mundler2024swtbench}, introduced by Mündler et al., is based on real-world Python issues, which evaluates performance through metrics like the Fail-to-Pass (F→P) rate and change coverage. 
Similarly, Ahmed et al. introduced TDD-Bench-Verified\cite{otter_ahmed_et_al_2025}, providing a more strictly filtered dataset and introducing 
a metric that assesses both the F$\rightarrow$P property and the test's coverage over code changes, offering a more fine-grained quality measure.

\textbf{No-Retrieval and Traditional Retrieval Methods.}
Early LLM-based approaches for generating BRTs are limited by their context retrieval strategies.
The no-retrieval approach, LIBRO\cite{libro_kang_yoon_yoo_2023}, utilizes a few-shot prompting paradigm, guiding the model with <issue report, test code> examples. 
Its performance is significantly limited by the complete absence of codebase-specific context in its prompts.
To address it, follow-up methods employ traditional information retrieval techniques. 
For instance, the ZeroShotPlus baseline in SWT-bench\cite{mundler2024swtbench} uses the BM25 algorithm to retrieve relevant code snippets from the repository. 
However, such term-frequency-based methods often fail to grasp the semantic meaning of the code, limiting the quality of the retrieved context.

\textbf{LLM-driven Retrieval Methods.}
With the advancement of model capabilities, more recent work leverages LLM for more intelligent retrieval. 
One retrieval method treats the LLM as an autonomous search agent.
For example, AEGIS \cite{aegis} constructs a dual-agent framework that includes a bug-related context summarization module and a finite-state-machine (FSM)-guided script generation module. 
Another retrieval approach
is hierarchical selection, which mimics the developer's workflow of first locating relevant files and then identifying specific functions within them. 
AssertFlip~\cite{khatib2025assertflipreproducingbugsinversion} adopts this method to locate faulty code for its ``pass-then-invert'' generation strategy, then treats all context needs as a single bug localization task.
The Otter series of methods (Otter, Otter++~\cite{otter_ahmed_et_al_2025}, e-Otter++~\cite{ahmed2025executionfeedbackdriventestgeneration}) refines this by using separate stages to retrieve code and test contexts. 
Otter introduces a ``self-reflective action planner'' to decide what to read and write, while e-Otter++ further enhances with execution feedback and generated patches. 
However, it applies the same retrieval strategy to both and relies on function name lists, overlooking complex function call relationships.

Different from existing retrieval methods for BRT generation, 
\appname designs differentiated retrieval strategies for relevant code and test cases, jointly considers textual semantics and function call structures when assessing relevance, and establishes a feedback iteration between retrieval and generation to optimize the context iteratively. 
By being aware of these correlations, 
\appname can retrieve more accurate and helpful context for bug reproduction, surpassing existing methods.
Moreover, our approach is generator-agnostic and can be easily integrated into more advanced BRT generation methods to enhance their overall performance.

\section{Conclusion}
This paper presents \appname, a novel approach for generating high-quality bug reproduction tests by retrieving context using a differentiated, two-stage strategy for code and test case retrieval.
By incorporating function call analysis to move beyond simple textual similarity, \appname can capture deeper behavioral relevance between tests.
Finally, \appname established a feedback loop from the generation phase to iteratively refine and improve the quality of the retrieved context.
Our comprehensive evaluation on the SWT-bench Lite benchmark demonstrates the superiority of our approach. 
When integrated with an LLM-based generator, \appname achieved a Fail-to-Pass rate of 42.0\%, which represents a significant 31.7\% relative improvement over the best prior retrieval methods.
Moreover, since \appname is generator-agnostic, future work can leverage it as a plug-and-play retrieval module, combining it with a more diverse range of generators to explore strategies for further improving the performance of BRT generation.

\section{Data Availability}
Our replication package, including the code and data, and our online appendix are available at \url{https://github.com/ZJU-CTAG/iCoRe}.

\begin{acks}
This research/project is supported by the National Natural Science Foundation of China (No.92582107) and the Hong Kong SAR Research Grant Council (General Research Fund Ref: 16206524).
\end{acks}

\bibliographystyle{ACM-Reference-Format}
\bibliography{ref.bib}

\newpage
\appendix

Our replication package, including the code and data, and our online appendix are available at \url{https://github.com/ZJU-CTAG/iCoRe}.

\section{Heuristic Selection}
\label{app:heuristic-alg}

Algorithm~\ref{alg:heuristic-ranking} provides the full pseudo-code of the heuristic-based selection mechanism described in Section~3.1.

\begin{algorithm}[hbp]
\caption{Heuristic-based Selection Algorithm for Candidate Nodes}
\label{alg:heuristic-ranking}
\begin{algorithmic}[1]
\REQUIRE A set of keyword retrieval results $\mathcal{R}$, where each keyword $k$ maps to a set of candidate nodes $N_k$.
\ENSURE The best node $\hat{n}_k$ for each keyword.

\STATE Initialize score for each node $n$: $S(n) \leftarrow 0$
\STATE Initialize score for each file $f$: $F(f) \leftarrow 0$

\FOR{each keyword $k$ and its corresponding node set $N_k \in \mathcal{R}$}
    \IF{$N_k$ is empty}
        \STATE continue
    \ENDIF
    \STATE Base weight $w_k \leftarrow 1 / |N_k|$
    \FOR{each node $n \in N_k$}
        \STATE file $f \leftarrow$ file containing node $n$
        \STATE $S(n) \leftarrow S(n) + w_k$ \hfill // Uniqueness weight
        \STATE $F(f) \leftarrow F(f) + w_k$ \hfill // File co-occurrence contribution
    \ENDFOR
\ENDFOR

\FOR{each keyword $k$ and its corresponding node set $N_k \in \mathcal{R}$}
    \FOR{each node $n \in N_k$}
        \STATE file $f \leftarrow$ file containing node $n$
        \STATE parent $p \leftarrow$ parent node of $n$ (e.g., class or module)
        \STATE Calculate total score: $T(n) \leftarrow S(n) + F(f) + S(p)$ \hfill // Weighted sum of three heuristics
    \ENDFOR
    \STATE Select the node with the highest score from $N_k$: $\hat{n}_k \leftarrow \arg\max_{n \in N_k} T(n)$
\ENDFOR

\RETURN The set of best nodes for all keywords $\{ \hat{n}_k \}$
\end{algorithmic}
\end{algorithm}

\section{Prompts for \appname}
\label{app:prompts}

\begin{promptbox}[Prompt for Keywords Extraction]\label{app:keyword_prompt}
You are a developer investigating a bug report. Your task is to extract critical code elements and relevant technical keywords from the report. These elements should be directly useful for reproducing the bug, or searching the codebase.\\
\\
Guidelines:  \\
1. Identify and list only the code-related elements that are essential for searching the codebase and understanding or reproducing the bug. These may include function names, class names, method names, variable names, file names, or other identifiers that directly contribute to debugging.\\
2. Exclude non-actionable or irrelevant terms, including: \\
- Generic words like ``feature'', ``error'', ``problem''.\\
- User-defined class names or model names that are created within the example bug report but are unlikely to exist in the actual codebase (e.g., \texttt{A}, \texttt{B}, \texttt{C} in a sample model definition).\\
- Any abstract or non-code terms that do not directly contribute to debugging.\\
3. Preserve the exact names or formats of the elements as written in the bug report. If an imported element is renamed using as, restore its original module path. For example, if the bug report mentions \texttt{import pandas as pd}, and pd.DataFrame is used in the code, extract it as \texttt{pandas.DataFrame}.\\
4. Prioritize the extracted elements by their importance for reproducing the bug:\\
- Elements that are most likely to be useful or necessary for understanding the bug should be ranked highest.\\
- Additionally, class names and function/method names should be ranked higher than code fragments.\\
\\
Output format: Provide the extracted code elements as a list in the following format:  \\
\texttt{[`ClassName', `ClassName.methodName', `functionName', `variableName']}
\\
For example, if the bug report mentions a class \texttt{MyClass} and methods \texttt{function1} and \texttt{function2} inside it, the output should be:  
\texttt{[`MyClass', `MyClass.function1', `MyClass.function2']}\\

\end{promptbox}

\begin{promptbox}[Prompt for Initial Test Retrieval]\label{app:initial_retrieve_prompt}
You are an assistant for a Python project, specifically helping to write test cases based on a bug report. Your tasks are as follows:\\
1. Find and recommend some appropriate test functions based on the bug report.\\
2. Rank the selected test functions in order of relevance, with the most relevant one first.\\
3. Output the name of test function and its file path. The result should contain at most \{topk\} test cases.\\
\#\#\# Available Tools:\\
- \texttt{list\_root()}: Lists all files and directories inside the root test folder of the project. You may call this function first.\\
- \texttt{list\_folder(path)}: Lists files and directories at the given path. \\
- \texttt{list\_classes\_and\_functions(file\_path)}: Lists all classes and functions in a given file.  \\
- \texttt{read\_function(file\_path, function\_name)}: Reads the source code of a specific function.  \\
You may need to call list\_folder multiple times, including on subdirectories, to explore the full directory structure and locate the appropriate test file. \\
To explore subdirectories, you need to use the relative path (e.g., tests/folder1/folder2). \\
If you find that the initial path you accessed is not suitable, you can access the directory, files, and functions multiple times to find the most appropriate ones.\\
The maximum number of steps is \{max\_steps\}.\\
\#\#\# Required Output Format:\\
Your output must follow this structure:\\
\texttt{[}\\
\texttt{["path/to/test\_file\_x.py", "test\_function\_m"],}\\
\texttt{...}\\
\texttt{["path/to/test\_file\_y.py", "test\_function\_n"]}\\
\texttt{]}\\

This list should contain at most \{topk\} entries, ranked from most to least relevant.\\

\end{promptbox}

\begin{promptbox}[Prompt for BRT Generation]\label{app:brt_gen_prompt}
You are an assistant responsible for expanding the test suite for our Python project. Your task is to write a test that FAILS on the current buggy code and PASSES after the bug is fixed, without needing any modifications.\\
Follow these instructions:\\
1. Understand the Bug Report and explicitly identify the following information:
Observed Behavior (OB): What is the actual, incorrect behavior produced by the current code?\\
Expected Behavior (EB): What is the correct behavior that the user expects?\\
\\
2. Design the Test Logic Based on the Expected Behavior\\
- Scenario A: If the Expected Behavior (EB) is the ``code should run successfully without errors''\\
Your test case should simply call the problematic code directly.\\
DO NOT wrap the code in try...except, pytest.raises, or assertRaises.\\
Reasoning: On the buggy code, the unexpected error (the OB) will be raised, causing the test to FAIL. After the fix, the code will run without error as expected (the EB), causing the test to PASS.\\
\\
- Scenario B: If the Expected Behavior (EB) is a ``specific error or warning should be raised''\\
Your test case MUST use with pytest.raises(SpecificError): (or an equivalent assertion) to wrap the problematic code.\\
Reasoning: On the buggy code, the specific error is not raised (the OB), so the assertion will fail, causing the test to FAIL. After the fix, the code raises the correct error as expected (the EB), and the test will PASS.\\
\\
3. Test Requirement\\
The test must be minimal and focused: Only reproduce the issue — do not add extra assertions that are unrelated to the bug.\\
\\
4. Use Provided Information Effectively\\
The input includes the following sections to assist in test creation:\\
<issue>: Contains the bug report. Focus on understanding the issue and its context.\\
<code>: Contains relevant code snippets. These can help you understand the project structure and usage of specific functions or classes.\\
Some code snippets may be irrelevant - use only what is necessary for constructing the test case. \\
If the bug report refers to a class or function that is a user-defined example, do not attempt to import such classes or functions. Instead, define a minimal version of the missing class or function within the test case, just enough to reproduce the bug.\\
<test>: Contains existing test cases that may relate to the bug. Analyze what each test does, evaluate its relevance to our bug, and determine whether it can be referenced to assist in generating a new test case. The test case you generate will be inserted into the same file as the related tests. Ensure the new test case reproduces the issue and aligns with the overall test suite structure.\\

\end{promptbox}

\end{document}